\documentclass[12pt]{article}

\usepackage{newtxtext,newtxmath}

\usepackage{graphicx}
\usepackage{amsfonts}

\usepackage{amssymb}
\usepackage{amsmath}
\usepackage[normalem]{ulem}
\usepackage{xcolor}
\usepackage[labelfont=bf, font={small,it}]{caption}
\usepackage{float}
\usepackage{subcaption}

\usepackage[letterpaper,margin=1in]{geometry}

\renewenvironment{abstract}
	{\quotation}
	{\endquotation}

\date{}

\makeatletter
\renewcommand{\fnum@figure}{\textbf{Figure \thefigure}}
\renewcommand{\fnum@table}{\textbf{Table \thetable}}
\makeatother

\usepackage[numbers]{natbib}

\usepackage{hyperref}

\newcommand{\squote}[1]{`#1'}

\def\papertitle{
	Propagation and preservation of AI-discovered problem-solving strategies in human culture
}

\title{\bfseries \boldmath \papertitle}

\author{
	Levin~Brinkmann$^{1,\ast,\dagger}$,
	Thomas~F.~Eisenmann$^{1,\ast,\dagger}$,
	Anne-Marie~Nussberger$^{1,\dagger}$,\\
        Maxime~Derex$^{2}$,
        Sara~Bonati$^{1}$,
        Valerii~Chirkov$^{1,3,4}$,
        Iyad~Rahwan$^{1,\ast}$\\
	\small$^{1}$Center for Humans {and} Machines, Max Planck Institute for Human Development, Berlin, Germany.\\
	\small$^{2}$Toulouse School of Economics, CNRS, Toulouse, France.\\
	\small$^{3}$Institute for Theoretical Biology, Department of Biology, Humboldt University Berlin, Berlin, Germany.\\
	\small$^{4}$Science of Intelligence, Research Cluster of Excellence, Berlin, Germany.
}

\begin{document} 
\maketitle

\vspace{-3.5em}
\begin{center}
\small
$^{\ast}$Corresponding authors. Email: brinkmann@mpib-berlin.mpg.de (L.B.);\\
eisenmann@mpib-berlin.mpg.de (T.F.E.); rahwan@mpib-berlin.mpg.de (I.R.)\\
$^{\dagger}$These authors contributed equally to this work.\\[6pt]
Author's version of the article published in \textit{Nature Communications} \textbf{17}, 8255 (2026).\\
\href{https://doi.org/10.1038/s41467-026-76113-2}{https://doi.org/10.1038/s41467-026-76113-2}
\end{center}

\begin{abstract} \bfseries \boldmath
Intelligent machines have the potential to uncover problem-solving strategies beyond human discovery. Emerging evidence from competitive gameplay, such as Go and chess, demonstrates that AI systems are evolving from mere tools to sources of cultural innovation adopted by humans. However, the conditions under which intelligent machines transition from tools to drivers of persistent cultural change remain unclear. We identify three key dimensions that modulate machine influence on human problem-solving: the discovered strategies must be non-trivial, learnable, and offer a clear advantage. Using a cultural transmission experiment, we demonstrate that when these conditions are met, machine-discovered strategies can be transmitted, understood, and preserved by human populations, leading to enduring cultural shifts. Conversely, using agent-based simulations, we show how machine influence is constrained in the absence of these conditions. These findings provide a framework for understanding how machines can persistently expand human cognitive skills and underscore the need to consider their broader implications for human cognition and cultural evolution.
\end{abstract}

\section{Introduction}
\noindent
Machines powered by Artificial Intelligence (AI) have the potential to discover strategies that are difficult for humans to conceive: their computational capacity exceeds human processing power and speed; their ability to process and share information in serial or parallel with high fidelity allows them to distribute problem-solving efficiently \cite{griffiths_understanding_2020}. AlphaGo, an AI system that defeated world champion Lee Sedol in the ancient game of Go, discovered superhuman strategies by playing millions of games against a copy of itself, and integrated these experiences into a complex strategy encoded in its neural network \cite{silver_mastering_2016}. Given the human-defined objective of winning the game, the algorithm used extensive search to identify a successful strategy that has been described as "alien", considering thousands of years of human Go-play \cite{sang-hun_googles_2016}. Recent evidence suggests that humans have significantly improved their own Go-play as a result of this innovation \cite{shin_superhuman_2023}, while more subtle shifts that appear less "alien" when viewed in historical context are also evident in Go opening moves \cite{beheim2025opening}. In chess, selected strategic concepts extracted from an AI system were applied by grandmasters after observing the AI system only on a few examples \cite{schut2025bridging}. The domain of gameplay remains, however, a rare example of human problem-solving strategies being shaped by machine-discovered strategies. This raises a question: Beyond being mere tools, under which conditions can machines persistently reshape the way we solve problems?

We address this question by integrating insights from the field of Cultural Evolution \cite{brinkmann_machine_2023}: culture changes through a process of variation, transmission, and selection \cite{boyd_culture_1985, mesoudi2011cultural}. Accordingly, the space of culturally adopted strategies is constrained by human discovery capacity, transmission fidelity, and the ability to perceive utility \cite{henrich2015secret,thompson_human_2021,lewis2012transmission}. We propose that machines’ capacity to discover novel strategies can expand these boundaries, contingent on three critical factors. \emph{Appropriate Discovery Difficulty:} Strategies must be non-trivial and extend beyond the range of variations typically discovered by humans. \emph{Low Transmission Difficulty:} Humans must be able to learn and effectively disseminate machine-introduced discoveries. \emph{Recognizable Selective Advantage:} Cultural adoption of machine discoveries hinges on their demonstrated success and the extent to which humans attend to them and endorse them. As such, our proposal integrates strands of literature that have remained largely separate: prior research highlighting the transformative power of selective social learning \cite{henrich_evolution_2001, kendal_social_2018, thompson_complex_2022}; observational evidence for the general possibility of machine-induced cultural shifts \cite{silver_mastering_2016, shin_superhuman_2023, schut2025bridging}; and conjectures about the complementary potential of human limitations and machine capacities \cite{griffiths_understanding_2020}.

\begin{figure}[!ht] 
  \centering
  \includegraphics[width=1\textwidth]{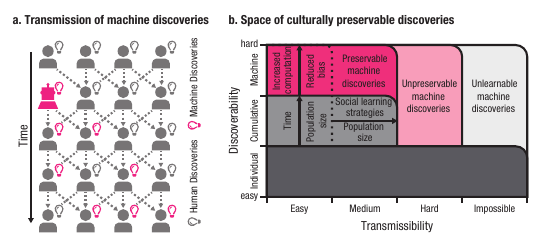}
\caption{\textbf{Machine-driven cultural shifts.} \textbf{A. Transmission of machine discoveries can lead to cultural shifts.} The multitude of potential transmission paths enables populations to preserve strategies that are difficult to transmit, fostering resilience against cultural loss. Machines, with their unique capabilities, may discover strategies that are improbable—or difficult to conceive—for humans. If such machine-discovered strategies are successfully transmitted and preserved within human populations, they could drive cultural shifts that outlast the presence of machines. \textbf{B. Space of culturally preservable strategies.} Human populations exploit strategies that are either easy to learn or sufficiently transmissible, forming a cultural frontier that has expanded with larger population sizes, time, and the development of social learning mechanisms (dark and medium gray). Machines, by virtue of increased exploratory capacity and reduced inertial bias, may discover strategies that are practically inaccessible to humans. We propose that some of these strategies may nonetheless remain preservable by humans (dark pink). Conversely, other machine-discovered strategies may be neither preservable nor even learnable by humans, potentially fostering a reliance on machines (light pink and light gray).}
\label{fig:concept}
\end{figure}

In more detail, for a new problem-solving strategy to become part of the human cultural repertoire, it first needs to be discovered. Human discovery is limited by computational capacities and pervasive cognitive biases \cite{simon_behavioral_1955, tversky_judgment_1974, griffiths_understanding_2020}. For example, human planning is generally confined to a few steps due to the combinatorial explosion of possible actions and outcomes \cite{simon_human_1971}. In adapting to this challenge, humans use a range of heuristics \cite{simon_behavioral_1955}, such as prioritizing immediate over long-term outcomes \cite{frederick_time_2002} or favoring familiar options \cite{goldstein_models_2002}. In contrast, machines surpass humans in computational capacity, allowing them to rapidly accumulate experiences that would be too costly in terms of time, computation, or risk for humans to access. This difference in resource constraints allows machines to explore much larger hypothesis and action spaces, reducing reliance on strong priors and heuristics that shape human search and enabling the discovery of strategies outside the region typically accessible to human exploration \cite{gershman2015computational}. This may lead machines to discover strategies that are unknown and potentially even counterintuitive to humans yet have superior performance (see Figure \ref{fig:concept}B, vertical axis). Experimental work has yet to investigate how computational heterogeneity and asymmetric exploration capacities shape discovery in human-machine societies, thereby extending prior work which typically considers one agent type \cite{thompson_complex_2022, derex_experimental_2013, acerbiLargeLanguageModels2023a}.

Discovering superior strategies is not sufficient to induce a cultural shift, however \cite{rahwan_analytical_2014}. The superior performance of the machine-discovered strategy must also be easy for humans to recognize as such \cite{henrich_evolution_2001, kendal_tradeoffs_2005, mcelreath_beyond_2008, morgan_evolutionary_2011}. Furthermore, humans must be able to socially learn the strategy and transmit it to others \cite{boyd_culture_1985}. The ability to select superior strategies and to transmit them in combination enables the population to maintain complex strategies that would be beyond individuals' reach \cite{thompson_complex_2022, miu_innovation_2018}. When these mechanisms apply to machine-discovered strategies as well, machines could become sources of new knowledge that shift human culture (see Figure \ref{fig:concept}B, horizontal axis). 

Even though intelligent machines are increasingly used in domains of human discovery, such as education and research \cite{luo_large_2024, sanchez-lengeling_inverse_2018, fortunato_science_2018}, in most cases these AI systems do not convey problem-solving strategies that are entirely alien to human culture. As such, they do not satisfy the above conditions but remain within the confines of existing human discoveries (see Figure \ref{fig:concept}B dark and medium gray). And even those AI systems that have discovered super-human knowledge, often remain – at present – tools in human hands rather than integrated into human cognitive frameworks. AlphaFold provides a remarkable example for this category: while it was inspired by the success of AlphaGo in emulating the intuition of human experts \cite{heaven_this_2022} and in doing so has achieved solving protein-folding problems at superhuman levels \cite{jumper_highly_2021, tunyasuvunakool_highly_2021}, its discoveries have not (yet) transpired into novel human intuitions, unlike in the case of AlphaGo and AlphaZero. As such, it presents an unlearnable and unpreservable machine discovery (see Figure \ref{fig:concept}B light gray).

\begin{figure}[!ht]
  \centering
  \includegraphics[width=1\textwidth]{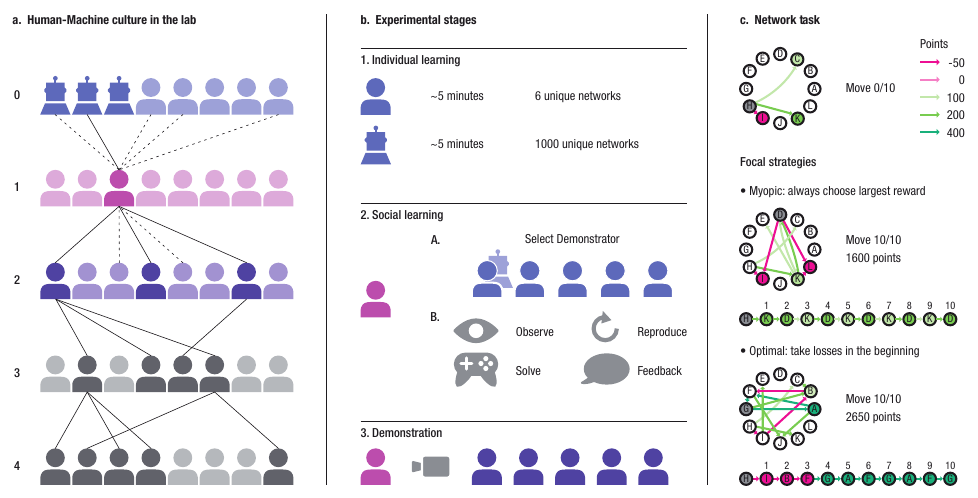}
\caption{\textbf{Experimental design and network task.} \textbf{A. Human-Machine Culture in the Lab.} Participants were divided into populations of five Generations (0-4) with eight individuals each, including 15 Human-Machine Populations (three machines in Generation 0) and 15 Human-Only Populations. Lines represent a subset of demonstrator-learner connections (dashed: potential, solid: selected) involving a focal participant (dark magenta) in Generation 1 and their cultural ancestor (dark blue) and descendants (dark violet and grey). \textbf{B. Experimental Stages.} The 'Individual Learning Phase' lasted five minutes, during which humans in Generation 0 managed to train on six networks, while machines managed to train on 1,000 networks. In the 'Social Learning Phase', Generations 1-4 were informed about the performance of 5 randomly drawn players from the previous generation (demonstrator candidates), from which they selected their preferred demonstrator. They then observed and replicated the demonstrator's solutions before solving the same network themselves and receiving feedback. In the 'Demonstration Phase,' individuals recorded their solutions for the next generation. \textbf{C. Network Task.} Participants collected points on a network, with new potential connections appearing sequentially. We illustrate the two focal strategies in our task: maximizing immediate gains (myopic), and incurring initial costs for greater future rewards (optimal).}
\label{fig:method}
\end{figure}

Here, we show that machines can persistently reshape human problem-solving culture when they discover strategies that humans typically fail to produce, yet can easily learn and transmit. To test this, we designed a behavioral task in which machines can produce counterintuitive solutions with a recognizable selective advantage. The task emulated a prototypical situation in which human exploration is constrained by the pervasive human tendency to avoid losses (i.e., myopic decision-making). Human participants (n = 1,155) were paid to solve a series of “reward networks” (Figure \ref{fig:method} C), which we developed as a modified version of an established paradigm \cite{huys_bonsai_2012, brinkmann_hybrid_2022}. In our task, networks consisted of nodes connected by links that were associated with rewards ranging from -50 to +400 points. Participants' goal was to choose 10 moves through the network that would maximize the number of points gained. Our networks were designed such that there was an optimal strategy for solving them: incurring early losses subsequently led to links associated with the maximum reward of +400 points, which were not accessible when avoiding early losses. Hence, the optimal strategy for solving our networks contradicts the myopic tendency of humans to avoid losses and prioritize immediate gains, even when accepting short-term losses could be beneficial in the long run. In line with prior literature, we expected that this myopic tendency would make the discovery of an optimal strategy in our task difficult for human players \cite{thompson_human_2021}. Meanwhile, we expected that a few minutes of training on our task would be sufficient for a deep-Q-learning algorithm (the "machine player") to successfully discover the optimal strategy through unbiased exploration at scale. To compare the performance of human players on their own against the performance of human players exposed to machine players, we randomly allocated participants into populations that involved either only eight humans (15 "Human-Only Populations") or five humans and three machine players in their initial generation (15 "Human-Machine Populations"). Each population consisted of five generations (numbered 0-4; Fig. \ref{fig:method} A), allowing us to observe differences between the two kinds of populations over time. After a series of six individual attempts at solving networks, participants in Generation 0 solved four networks during the "Demonstration Phase", knowing that their moves would be recorded for future generations to learn from. After two individual trials to familiarize themselves with the tasks, participants from Generations 1-4 could choose a "demonstrator" from the previous generation to learn from, based on the demonstrators' average performance. They did not know the demonstrators' identity as a human or machine, but were informed that machine players were a possibility. They then observed animations of the demonstrator’s moves, replicated them, and subsequently solved the same networks before receiving final feedback, with this procedure repeated across four networks. Only after this Social Learning Phase did participants from Generations 1-4 demonstrate their moves for the next generation. All networks solved by the participants throughout the experiment were unique, requiring them to discover or learn a generalizable strategy rather than optimizing a specific solution. 

In a second exploratory step, we sought to follow up on our experimental results with an agent-based simulation. Removing the feasibility constraints that apply to large-scale experiments with human participants, the simulation allowed us to systematically vary discovery and transmission difficulty, and to manipulate the extent to which social learning is selective \cite{kendal_social_2018}. In parallel to the experimental set-up, we organized agents into five generations, each consisting of eight agents. Agents could discover strategies through exploration or adopt strategies from the previous generation. Abstracting away from the task we used in the experiment, agents explored a simplified strategy space comprising the strategies: \textit{myopic}, and \textit{optimal}.  For “machine” agents, we independently manipulated exploration capacity and bias to investigate their impact on the strategies adopted within hybrid populations across subsequent generations.

\section{Results}
\subsection{Humans Preserve the Optimal Strategy Discovered by Machines}

\begin{figure}[!ht]
  \centering
  \includegraphics[width=1\textwidth]{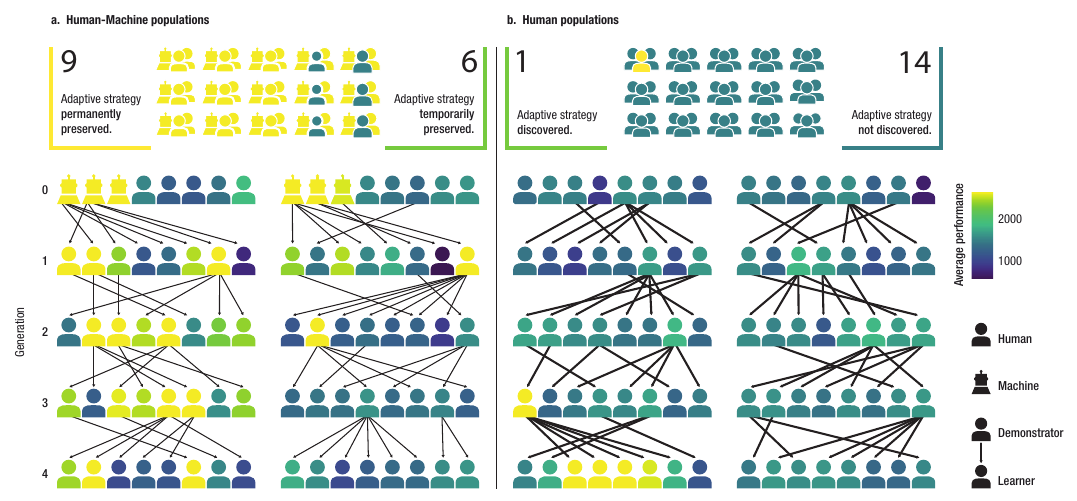}
\caption{\textbf{Four prototypical populations}. We categorized populations based on the average scores of their top performers across generations. Scores above 2000 points indicate adoption of the optimal strategy. Among 15 Human-Machine Populations (panel A), 9 consistently exceeded this threshold, while the remaining 6 surpassed it temporarily in at least one generation following the introduction of machines. In contrast, only 1 of 15 Human-Only Populations (panel B) temporarily achieved scores indicative of the optimal strategy, while all the other Human-Only Populations did not. We illustrate one prototypical population for each case, with colors reflecting performance scores. Arrows represent demonstrator-learner relationships.}
\label{figure2}
\end{figure}

In our reward network task, achieving more than 2000 points was only possible by following the optimal strategy of incurring early losses to reach later links associated with the maximum reward of +400 points per move. Generally, human players did not find this optimal strategy on their own: Among the human populations, only one out of 600 participants \textit{discovered} the optimal strategy over the course of the experiment, which was subsequently learned by four other human players. While the deep Q-learning agent, through extensive exploration over 20{,}000 epochs across 1{,}000 networks, discovered and reliably reproduced the optimal strategy within less than five minutes (Supplementary Fig.~\ref{fig:algorithm}), human participants, by contrast, exhibited a systematic avoidance of loss-incurring trajectories: paths involving three losses were under-sampled relative to a matched random baseline ($\mathrm{LOR}=-1.22$, $\mathrm{CI}_{95\%}=[-1.38,-1.06]$), reducing the likelihood of discovering the optimal strategy (see also Supplementary Text). Despite the rarity of individual human discovery, the optimal strategy readily spread from machines to humans in the human--machine populations and was preserved beyond the presence of the machines across multiple generations: In 9 out of 15 Human-Machine Populations, human players learned this strategy in each of their generations. Correspondingly, we classified these populations as \textit{permanently preserving} the machine discovery. In the remaining 6 populations, at least one human player in the generation immediately following the machine achieved more than 2000 points; but this discovery was only \textit{temporarily preserved} as it did not transmit to later generations. Figure \ref{figure2} illustrates prototypical cases for each of these qualitative categories. To summarise, the majority of Human-Machine Populations preserved the optimal strategy until the final generation in our experiment, while 14 out of 15 human populations did \textit{not discover} the optimal strategy at all. 


Preservation of the optimal strategy was supported by a bias toward selecting the demonstrator with the highest performance. Overall, $86.7\%$ ($\mathrm{CI}_{95\%}=[84.5, 88.7]$) of all participants chose the highest-scoring demonstrator. Consequently, regarding the cultural lineages emerging from demonstrator–learner relationships within the Human-Machine Populations (see Figure \ref{figure2}), we found that only machines had long-lasting impact on the cultural dynamics operating across generations: In the first socially learning generation of the Human-Machine Populations, most players (90\%) selected a machine as their demonstrator, and by the last generation, every player was a descendant of a machine player. In contrast, none of the human players in Generation 0 of the Human-Machine Populations initiated a cultural lineage that persisted until the last generation. 

\subsection{Humans Benefit from the Machine Discovery}
\begin{figure}[h]
  \centering
  \includegraphics[width=1\textwidth]{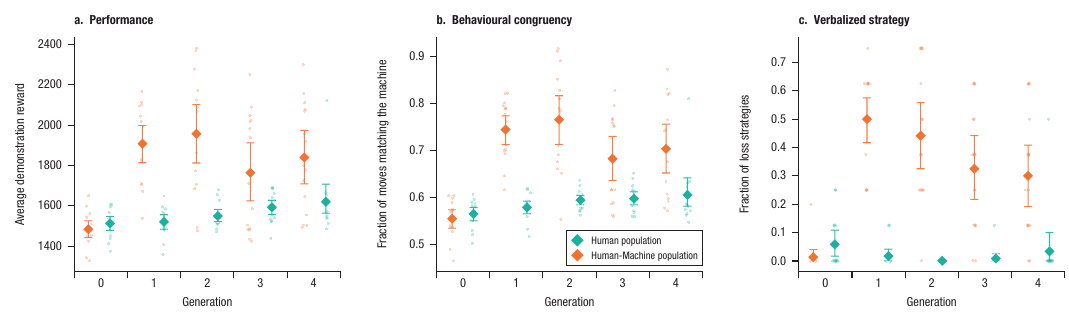}
\caption{\textbf{Evolution of Task Performance, Behavioural Congruency, and Strategy Description of Human Participants}. Individual populations are represented as dots, with the grand average for each condition indicated by a diamond symbol. Human-Only Populations are indicated in green, Human-Machine Populations in orange. Performance (Panel A) refers to the average number of points earned during the Demonstration Phase. Behavioural Congruency (Panel B) indicates the proportion of human player moves that are congruent with the majority behavior of the three trained machines. Verbalised Strategy (Panel C) shows the fraction of human players who articulated a strategy referring to a deliberate acceptance of losses after the Social Learning Phase. Data are presented as mean values. Error bars indicate 95\% bootstrap confidence intervals (CI) around the mean, computed over population-level aggregates. $n = 15$ populations per condition, each comprising 8 participants per generation. Each dot represents one population. The Human-Only condition serves as the control group. Participant-level means are reported in Supplementary Figure~\ref{fig:metrics_individual_level}.}
\label{figure3}
\end{figure}

Panel A in Figure \ref{figure3} shows the average reward gained in the task by generation and population. Descriptively, while the two types of populations started out with similar descriptive averages in Generation 0, in Generations 1-4 Human-Machine Populations consistently earned higher rewards. Our preregistered linear mixed-effects model on these latter generations supported that this difference was significant ($\beta = -309.2$, \textit{SE} = $57.0$, $\text{CI}_{95\%}$ = [$-420.7$, $-196.6$], $P < 0.001$). This model corresponds to model 1a for reward in the preregistration and includes an interaction between condition and generation (like model 2b). Further models confirmed that Human-Machine Populations performed better than Human-Only Populations both in Generation 1, which could learn directly from machine players (reward model 2a: $\beta = -404.6$, \textit{SE} = $59.4$, $\text{CI}_{95\%}$ = [$-513.5$, $-293.0$], $P < 0.001$), and in  Generation 4, where influence from machines was remote by four iterations (reward model 1b: $\beta = -229.1$, \textit{SE} = $82.4$, $\text{CI}_{95\%}$ = [$-407.4$, $-72.3$], $P = 0.010$).

\subsection{Humans are Behaviourally Congruent with the Machines}

Panel B in Figure \ref{figure3} shows the proportion of human moves congruent with machine-generated moves, summarized by generation and population. Qualitatively, the results are similar to the results for rewards: no statistically significant difference between conditions in Generation 0, but Human-Machine Populations are consistently more behaviourally congruent in Generations 1 to 4. A logistic mixed-effects model on these generations shows that Human-Machine Populations were more congruent than human populations (congruency model 1a: $\beta = -0.87$, \textit{SE} = $0.14$, $\text{CI}_{95\%}$ = [$-1.12$, $-0.61$], $P < 0.001$). Again, this was also the case within Generation 1, which learned from machine players (congruency model 2a: $\beta = -1.13$, \textit{SE} = $0.13$, $\text{CI}_{95\%}$ = [$-1.41$, $-0.88$], $P < 0.001$), and Generation 4 (congruency model 1b: $\beta = -0.64$, \textit{SE} = $0.20$, $\text{CI}_{95\%}$ = [$-1.02$, $-0.25$], $P < 0.001$). Notably, the individual who independently discovered the optimal strategy, as well as subsequent participants, likewise exhibited elevated congruency with machine behavior (see Supplementary Figure \ref{fig:populations_alignment}).

\subsection{Humans Internalize the Machine Discovery}

During the Demonstration Phase of our experiment, players were asked to provide a written strategy of how they approached the task. In contrast to the demonstration itself, these written strategies were not transmitted to the next generation, and players were aware of this. Panel C in Figure \ref{figure3} shows the proportion of human players mentioning the optimal strategy (i.e. the "loss strategy") in their written statements, as coded independently by three human annotators. In the Human-Only Populations, we observe practically no mention of the strategy, with proportions close to 0 throughout all generations; the exception to this is the single population that discovered the optimal strategy in Generation 3, resulting in 50\% of players referring to it in Generation 4. In contrast, participants in the Human-Machine Populations consistently mentioned the optimal strategy explicitly, with proportions of about 30\% to 50\% per generation. A logistic mixed-effects model on Generations 1 to 4 shows that this difference is significant (model 2b: $\beta = -4.18$, \textit{SE} = $0.62$, $\text{CI}_{95\%}$ = [$-5.49$, $-3.04$], $P < 0.001$).


\subsection{Simulation Corroborates Conditions for Cultural Shift}
\begin{figure}[ht] \centering \includegraphics[width=1\linewidth]{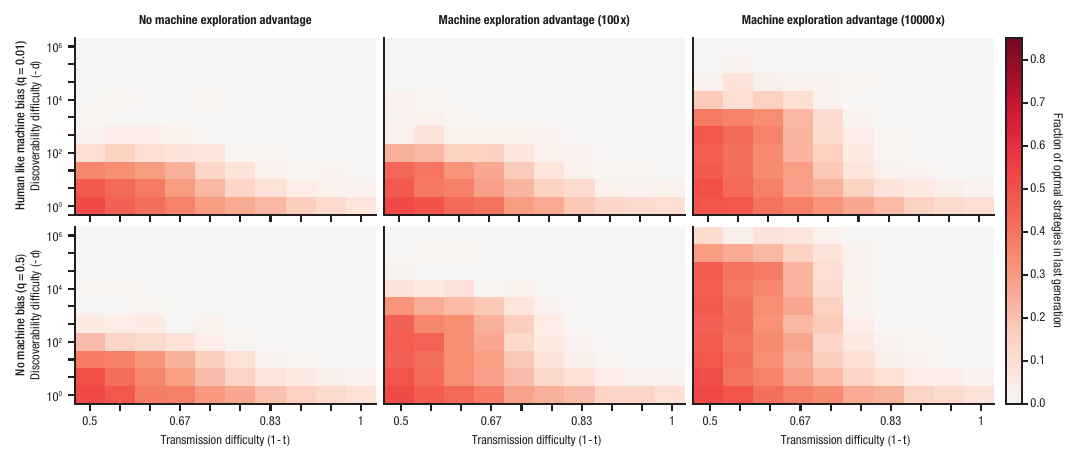} \caption{
\textbf{Agent-Based Simulation Highlights Conditions for Cultural Shifts.} We developed an agent-based simulation replicating our multigenerational social learning experiment (20 generations of 20 agents), systematically varying the difficulty of discovering and transmitting the optimal strategy. Five “machine” agents were introduced in the early generations, and we manipulated their exploratory capacity and bias toward myopic solutions relative to “human” agents. The heatmap shows the proportion of agents in the final generation (in the absence of machines) that adopt the optimal strategy. The upper-left panel is equivalent to a scenario without machines. Increased exploration and reduced bias in machine agents enable the discovery of strategies beyond human reach, which are subsequently adopted and maintained by human agents. Color scale represents the fraction of optimal strategies in the last generation, ranging from 0 (light) to 0.8 (dark).} \label{fig.abm} \end{figure}

We devised an agent-based simulation to systematically examine how variation in discovery and transmission difficulty shapes the cultural impact of machines operating at superhuman capacity. Agents attempted one of two strategies, myopic or optimal, over a finite number of trials $K$, and discovered the corresponding strategies through a binomial process with success rate $d$. A bias parameter $q$ controlled the likelihood of attempting the optimal rather than the myopic strategy. We simulated 20 generations of 20 agents each, in which agents could also acquire strategies through social learning from the previous generation with success rate $t$.

We first confirmed that, in homogeneous populations and under high social learning success rates $\lambda = 0.9$, discovery of strategies of higher difficulty is constrained by bias $q$, limited exploratory capacity $K$, and the number of generations elapsed, $G$ (see Supplementary Figure \ref{fig.bias_discoverability}). We then focused on scenarios with heterogeneous exploration capacity and bias, in which the majority of agents (i.e., human agents) are limited in their exploration capacity $k_{\text{m}} = 10$ and exhibit a bias $q_{\text{m}} = 0.01$ toward myopic solutions. We systematically manipulated the exploratory capacity and bias of five machine agents added to the first generation and examined the resulting adoption of optimal strategy by human agents in the final generation (Figure~\ref{fig.abm}). Increasing the exploratory capacity $K_{\text{m}} \in [1000, 100000]$ of the machine agents and neutralizing their bias toward myopic strategies $q_{\text{m}} = 0.5$ both increased the likelihood of discovering the optimal strategy. Consequently, introducing such machine agents enabled populations to adopt optimal strategies even under high discovery difficulty and thus expanded the horizon of obtainable strategies within a population. 

The simulations further show that when optimal strategies are difficult to transmit, they fail to spread and persist across generations, thereby constraining machine-induced cultural shifts. This effect is amplified when agents are less able to identify and select successful models for social learning, either because they are insufficiently sensitive to performance differences (higher $\tau$) or because realized rewards are highly noisy (higher $\epsilon$). Under these conditions, optimal strategies neither diffuse nor persist within populations (see Supplementary Figures~\ref{fig.abm_social_learning_selection} and~\ref{fig.abm_reward_uncertainty}).
 
\section{Discussion}

Our work demonstrates that AI discoveries can persistently reshape human problem-solving strategies when they are calibrated in difficulty, transmissibility, and selective advantage. In our experiment, exploration at a capacity several orders of magnitude greater allowed machine players to discover a superior strategy that offset human cognitive biases toward myopic and suboptimal solutions and, as such, was effectively beyond human reach within the constraints imposed by the experiment. This does not mean that the optimal strategy was impossible to discover for human participants at all, but limitations in time and cognitive biases \cite{griffiths_understanding_2020} made it exceedingly rare. Yet, the machine-discovered strategies were still learnable for human players, who did not merely replicate them but also applied them to new variations of the task. Further signifying the integration of the counterintuitive problem-solving strategy into their cognitive framework, many human players explicitly articulated the strategy in writing. The machine discovery's selective advantage translated into human players favoring machine over human ancestors, producing persistent cultural lineages where every human player in the last generation was ultimately a descendant of a machine player in the first generation. Consequently, the optimal strategy was adopted by the majority of populations with machine ancestors, whereas human populations predominantly employed the myopic strategy, indicating the capacity of machines to act as drivers of cultural change.

Human culture supports the cumulative accumulation of knowledge and skills across generations, enabling individuals to benefit from collective experience that far exceeds what any single person could acquire in isolation \cite{henrich2015secret,mesoudi2011cultural}. Laboratory studies explicitly compressing generational timescales into hours or minutes show that social transmission can overcome time constraints and the costs of individual exploration \cite{caldwell2008studying,derex2024trade}. Building on this literature, the present work suggests that machines may function as novel sources of cumulative cultural input. By learning from machines, humans may adopt and transmit hard-to-conceive optimal strategies, thereby expanding their cognitive repertoire without incurring the cognitive costs and time typically associated with discovering such strategies. Our results apply particularly to scenarios in which human exploration is constrained, while machine exploration remains scalable. For instance, chemical recipes can be explored at high throughput and scale through automation \cite{gromski_universal_2020}. In our experiment, the state space was small enough to permit brute-force exploration, and the primary role of Q-learning was to extract a generalizable, model-free strategy from this experience. In many real-world domains, however, solution spaces are intractable, and exploration is resource-intensive. Future research should examine whether the extensive world knowledge embedded in large generative models might similarly facilitate previously inaccessible discovery processes when both humans and machines operate under comparable data constraints.

Despite most populations maintaining the algorithmically discovered optimal strategy, the results show that even selectively advantageous strategies remain vulnerable to stochastic loss under finite populations and transmission noise. Prior work in cultural evolution indicates that large effective population sizes can buffer against the loss of complex discoveries \cite{henrich_demography_2004, derex_experimental_2013}, suggesting that machine-induced cultural shifts may be amplified when large groups interact with machine-generated strategies and transmission is the main constraint. For instance, the widespread adoption of generative AI in educational, professional, and creative contexts exposes large populations to machine-mediated strategies. Larger groups increase the likelihood that such strategies are learned, as skilled individuals transmit them to others through repeated demonstrations \cite{derex_experimental_2013}, facilitating the accumulation of a critical mass necessary for tipping points in cultural adoption \cite{centola_experimental_2018}. By contrast, in smaller populations, machines’ advantages in discovery can be particularly pronounced, given the limited collective exploration of such human groups. Likewise, emerging evidence suggests that machines’ ability to rapidly identify adaptive solutions within a population can enhance collective efficacy \cite{ueshima_simple_2024}.

Yet another factor that could broaden the scope of machine-induced cultural shifts is pedagogical behavior. Teaching is one of the primary mechanisms through which humans transmit knowledge and skills \cite{boyd_culture_1985, csibra_social_2006}, and experiments have shown that the value of teaching increases with the complexity of strategies \cite{lucas_value_2020}. Notably, discovery and transmission difficulty have also been identified as key determinants in the evolution of teaching \cite{fogarty_evolution_2011}, underscoring that machine-induced cultural shifts may occur within a relatively narrow range of conditions. This line of work does, however, also point towards another possible pathway that could enhance the efficacy of machine-induced cultural shift: more explainable AI systems could provide both teachers and learners with insights that facilitate the adoption of machine-discovered strategies, and enhance their scope through more effective human-machine interactions \cite{malone_collective_2010}. In addition, recent work in chess suggests that testing for AI-to-AI teachability could be a promising way to identify strategic machine concepts that can be effectively transferred to humans \cite{schut2025bridging}. At the same time, pedagogical features in AI systems may not always be necessary or desirable, particularly when humans use these systems primarily as tools. As AI systems are increasingly trained on one another’s outputs, they may transmit adaptive strategies in the form of cultural artifacts that are opaque to humans, potentially leading to a bifurcation of human and machine cultures \cite{brinkmann_machine_2023}.

A major question arising from these results concerns the scope of machine-induced cultural shifts. In pilot experiments, we found a key challenge in achieving the right balance between discovery difficulty and transmissibility (see Supplementary Text): At one extreme, simple strategies were readily discovered by humans without machine assistance, limiting their capacity to induce meaningful shifts. At the other extreme, strategies requiring lengthy and precise sequences of actions were too complex for human participants to learn and pass on. It is remarkable that the optimal strategy in our final design was discovered by none of the 195 human participants in the first generation (in deviation from previous work such as \cite{thompson_complex_2022}, where the discovery rate in the first generation was still at 13\%). Nonetheless, in our setting, this optimal strategy was still learnable and preservable once it was discovered. As a controlled online experiment, our design necessarily reflects stylized conditions, with individual learning constrained to brief exploration periods that may differ from real-world settings. However, analyses of participants in generation 0 with greater learning opportunities provide no statistically significant evidence that extended exploration budgets attenuate loss-related biases (see Supplementary Text).

Theoretical frameworks about cumulative cultural evolution posit that culture evolves through the iterative refinement of adaptive solutions and strategies \cite{mesoudi_what_2018}, implying that certain outcomes may remain undiscoverable without intermediate adaptive steps. The results presented in this work suggest that the scale of exploration alone can enable machines to discover strategies that humans fail to uncover, but are able to learn. They also indicate an interplay between exploration scale and bias, or cultural entrenchment, that constrains human exploration. Building on this insight, future studies could explore how the enhanced exploratory depth and breadth enabled by machines might help overcome structural limitations, such as entrenched path dependencies and biased expectations shaped by a historically conditioned epistemic horizon  \cite{clauset_data-driven_2017, fortunato_science_2018}. Supporting this conjecture, recent work has highlighted machines' potential to identify novel strategies for fostering cooperation that humans have proven unlikely to consider \cite{mckee_scaffolding_2023}. Even when it comes to changing notoriously entrenched conspiracy beliefs, intelligent machines (here, LLMs) are emerging as potent agents of persistent learning among humans \cite{costello_durably_2024}. But these contexts also point to the potential challenge that machine-discovered strategies might sometimes be undesirable from a human perspective and not be explored for good reasons, such as ethical concerns or social acceptability. Relatedly, human adoption and preservation of machine-discovered strategies may raise ethical concerns about human autonomy and creativity as much as it might bear epistemic risks that could undermine human knowledge \cite{messeri_artificial_2024}. In other contexts, AI aversion or other (oftentimes variable) perceptions of machine-derived knowledge \cite{dietvorstAlgorithmAversionPeople2015, loggAlgorithmAppreciationPeople2019, brinkmann_hybrid_2022} might influence adoption rates. However, as it becomes increasingly difficult to dissociate human- from machine-derived cultural artifacts \cite{yakura2024empirical}, it is all the more crucial to understand how machines might expand the frontiers of human thought and enable the emergence of culturally significant, hard-to-conceive strategies.
\section{Methods}
This study was approved by the Ethics Committee of the Max Planck Institute for Human Development (protocol A 2022-24). All participants provided informed consent prior to participation.

\subsection{Participant Recruitment, Exclusion and Compensation} 

Participants (N = 1,155) were recruited through the online platform Prolific \cite{palan_prolificacsubject_2018}, while screening for individuals residing in the United States and the United Kingdom, being proficient in English, and maintaining an approval rating of at least 95\%. Upon successful completion of the study, participants received a fixed payment of \textsterling{2.25}, complemented by performance-dependent bonus compensation payments averaging \textsterling{0.70}. The median time for completing the study was 13:45 minutes. The average hourly compensation rate was \textsterling{9.82}.

Participants ranged in age from 18 to 92 years (M = 39.67, SD = 12.28). Of those with available demographic data (n = 1,272), 56.4\% identified as male, 43.2\% as female, and 0.3\% preferred not to say. The majority resided in the United Kingdom (72.9\%) versus the United States (27.1\%). Sex and gender were not analysed as variables in this study as they were not relevant to the research question.

As preregistered, participants were excluded based on two criteria: 1) Participants who took more than 25 minutes, deemed unreasonably long; 2) Participants with missing moves in any of their demonstration trials, as subsequent generations could not replicate incomplete solutions. These criteria also implicitly served as an attention check. Of the 1,303 participants who completed the experiment, 148 were excluded under these conditions. Their spots were then filled by newly recruited participants. No other exclusions were made. The investigators were not blinded to allocation during outcome assessment, as the analysis followed a preregistered plan.

Participants who passed the exclusion criteria were eligible for bonus compensation. Participants' compensation depended in part on their own performance in the Demonstration Phase, and in part on the performance of people that learned from them. To ensure a fair chance of gaining a bonus payment independently of the condition a participant was sampled in, we ranked these scores separately per generation and condition. The final bonus was then computed as the percentile score rank multiplied by \textsterling 1.40.

\subsection{Preregistration}
The study was preregistered on asPredicted.org on 15 February 2024 (\url{https://aspredicted.org/d765-77fm.pdf}), including a detailed description of predictions, measures, conditions, analyses, exclusion criteria and sample size. No deviations from the preregistered analysis plan were made. In particular, we preregistered the following predictions:

\begin{enumerate}
    \item \textbf{Main predictions}
    \begin{enumerate}
        \item \textit{General benefit of AI}. Overall, participants from generation 1 onwards in AI trees will outperform participants in human trees.
        \item \textit{Long-lasting benefit of AI}. The subset of participants in the final generation of AI trees will outperform the final generation of the human trees.
    \end{enumerate}
    \item \textbf{Secondary predictions}
    \begin{enumerate}
        \item \textit{Transmission from AI to humans}. The subset of participants in generation 1 of the AI trees will outperform generation 1 in the human trees.
        \item \textit{Explicit recognition of counterintuitive strategy}. Overall, participants from generation 1 onwards in the AI trees will describe the counterintuitive strategy more frequently in their written strategies than those in the human trees.
    \end{enumerate}
\end{enumerate}

\subsection{Network Task}

We adapted the Reward Network task \cite{huys_bonsai_2012, brinkmann_hybrid_2022} consisting of a network \(g\) with 12 nodes and 30 edges, where each edge is associated with a reward \(r \in [-50, 0, 100, 200, 400]\) (see figure \ref{fig:method} panel C). Hidden to participants, each node was associated with a level \(l \in [0, 1, 2, 3]\). Constraints in the network generation ensured that any path connecting a node of level 0 to a node of level 3 included at least three edges with negative rewards. The largest reward of 400 points was accessible only from level 3. These two constraints required participants to accept at least three losses to access the highest rewards. Transitions between nodes of level 0 contained the second-highest reward of 200 points, making myopic behavior a reasonable but generally suboptimal strategy.

Participants began the game at a predetermined starting node (e.g. node B in figure \ref{fig:method} panel C). Initially, only the outgoing edges from this node were revealed. As participants traversed the network, they sequentially uncovered it and the outgoing edges of each visited node were displayed. Each participant would be playing on, in total, 10 different networks drawn randomly from the pool of experimental networks. The assignment of networks to individual participant trials ensured that no participant experienced the same network twice.

We generated three sets of networks (training, validation, and experiment) with 1,000 networks each. The first two sets were used only during the training of the neural policy, while the last set was used exclusively during the experiment. After generation, we computed scores for a myopic rule-based strategy and a loss-seeking rule-based strategy. The myopic strategy always selects the move with the immediate highest reward, while the loss-seeking strategy prioritizes edges with a -50 reward over all others. We excluded networks from all three sets where the myopic strategy outperformed the loss-seeking strategy, which accounted for approximately 9\% of all generated networks.

\subsection{Populations and Social Learning Opportunities}

The experiment consisted of 30 populations, each with 5 generations and 8 players per generation (see figure \ref{fig:method}.A). Half of the populations were assigned to the "Human-Machine" condition, and the other half to the "Human-Only" condition. In the "Human-Machine" condition, machines took three player positions in the first generation. Starting from generation 1, each participant could choose one of five possible demonstrators drawn from the eight positions in the previous generation. Participants were dynamically assigned to available positions in the populations, prioritizing the completion of existing populations before adding participants to new ones. Positions for later generations became available only after all potential demonstrators had completed the experiment.

\subsection{Experimental Flow}
\label{exp_flow}

Panel B in Figure \ref{fig:method} depicts the main phases of the experiment. A static version mimicking the experience for participants in later generations is available here: 

\vspace{5pt}
\url{https://center-for-humans-and-machines.github.io/reward-network-iii}. 
\vspace{5pt}

\noindent Below, we summarize the main phases of our experiment:

\begin{enumerate}
  \item \textbf{Introduction Phase}: Participants learned about the study's purpose, procedures, data handling, estimated time, and financial compensation. They then completed a tutorial guiding them through the task and interface.
  \item \textbf{Individual Learning Phase}: Participants solved the task individually. Generation 0 worked through 6 networks, while later generations worked through only 2 networks (given they encountered four more in the Social Learning Phase, see below) to familiarize themselves with the task. See also Figure \ref{fig:self_practise}.
  \item \textbf{Social Learning Phase (only from generation 1}: Participants selected a demonstrator (see Figure \ref{fig:select_teacher}). They could base their decision on demonstrators' average scores from the Demonstration Phase (they were also informed about their own average score as a reference point). They then went through four networks by observing a replay of the demonstrator’s solution (see Figure \ref{fig:observation}), repeating that solution (see Figure \ref{fig:repeat}), and trying to solve the same network on their own. Replay delays between moves were equalized to eliminate timing differences between human and machine-generated demonstrations. Last, they received feedback comparing their solution with the demonstration.
  \item \textbf{Demonstration Phase}: Participants independently solved four networks. The solutions entered were used for the Social Learning Phase of the next generation for those selected as a demonstrator. 
\end{enumerate}

Participants also documented their personal strategy in written form, before and after the Social Learning Phase. These strategies were collected solely for later analysis and were not presented to subsequent generations.

\subsection{Neural Policy and its Training}

We used a deep Q-learning-based reinforcement learning approach to train a neural policy for navigating and solving reward network tasks, as outlined by \cite{mnih_human-level_2015}. The neural policy featured a gated recurrent unit (GRU) \cite{chung_empirical_2014}, flanked by two linear layers with ReLU activation, each containing 15 hidden units. The input \(Q(o_t)\) consisted of a one-hot-encoded reward of each target node in a vector of shape \({Nodes \times Rewards}\), while the output represented the Q-value of each target node, with unreachable nodes set to negative infinity. Thus, similar to the initial experience of human participants, the algorithm could only process the immediate outgoing edges from the current node without engaging in any explicit planning.

The neural network approximated Q-values \( Q(o, a; \theta) \) where \( \theta \) represented the network parameters. These parameters were optimized by minimizing the loss function:

\begin{equation} L(\theta) = \mathbb{E}\left[\left(y_t - Q(o, a; \theta)\right)^2\right] \text{,} \end{equation}

with the target value \( y_t \) defined by the Bellman equation:

\begin{equation} y_t = r + \gamma \max_{a'} Q(o', a'; \theta^-) \text{,} \end{equation}

and \( \gamma = 0.99 \) as the discount rate. The target policy weights \( \theta^- \) were only updated every 200 steps to stabilize training.

An epsilon-greedy strategy guided training with \( \epsilon \) starting at 1 and reducing according to \( \epsilon_e = 0.99^{\lfloor e/1,000 \rfloor} \), reaching a minimum of 0.01. Trajectories were stored in a replay buffer of 500 episodes, and a batch of 16 episodes was sampled for updating the policy. The policy was optimized using the Adam optimizer with an initial learning rate of \( 1 \times 10^{-3} \), adjusted downwards by a factor of 0.8 every 2,000 episodes. The agent underwent 20,000 episodes of training, with evaluations every 100th episode using 1,000 test networks from a hold-out set. Training took approximately 5 minutes on an RTX 5,000 GPU. The neural policy was trained using Python 3.10 and PyTorch 2.1.1. Complete package versions are listed in the code repository. We depict the average performance on a hold-out test set in Figure \ref{fig:algorithm}, illustrating that the policies initially discovered the myopic strategy before eventually settling on the optimal strategy.

\subsection{Measures}

\paragraph{Task Performance} We measured task performance by calculating the total score for each trial, which involved summing the rewards from the 10 moves. The maximum achievable reward was 2,650, which corresponded to taking three losses of 50 points each and then gaining 400 points seven times. In contrast, the maximum score attainable with a myopic strategy was 2,000, yielded by ten gains of 200 points each. The exact scores achievable with these two strategies varied between networks.

\paragraph{Behavioural Congruency} We assessed human--machine behavioural congruency by comparing the actions taken each human $l$, denoted as \( a^H_{l,t'} \), with those predicted by an ensemble of three machines. For each machine \( k \in \{1,2,3\} \), the predicted action was defined as  
\begin{equation}
a^{M_k}_{l,t'} = \arg\max_a Q_k(a^H_{l,0}, \ldots, a^H_{l,t'-1}) \text{,}
\end{equation}
based on the trajectory of moves \( a^H_{lt} \) selected by the human participant. The ensemble prediction was then determined by majority vote,  
\begin{equation}
a^{M}_{l,t'} = \mathrm{mode}\big(a^{M_1}_{l,t'}, a^{M_2}_{l,t'}, a^{M_3}_{l,t'}\big) \text{.}
\end{equation}

A match \( a^H_{l,t'} = a^{M}_{l,t'} \), where the ensemble’s action corresponded with the human’s action, was encoded as `1'; a mismatch as `0'. This method required processing the entire trajectory to ensure that the recurrent neural units of the neural policies could utilize accumulated historical information to influence their decisions.
\paragraph{Written Strategies} 

To measure the explicit recognition of the loss strategy, we manually coded all 2,310 strategy descriptions (covering both time steps for every participant) with the aid of three trained human annotators. The annotators familiarized themselves with the task and independently coded the written strategies using a binary variable to indicate the presence of the loss strategy. Before coding the full experimental data, the annotators calibrated their ratings by coding a small independent dataset (n = 76) from two previous pilot studies. After individual coding, interrater agreement was high (Light's $\kappa = 0.83$). They then discussed any discrepancies before proceeding to code the full dataset. Annotators were blind to participant conditions during both calibration and full data coding, and data was randomly shuffled prior to coding to prevent any bias linked to participant order. For the full dataset, interrater agreement was excellent (Light's $\kappa = 0.92$).

\subsection{Statistical procedure}

All analyses were performed using the \textit{lme4} package (version 1.1-35.1; \cite{bates_fitting_2015}) in R (version 4.3.2; R Core Team), and the structure and procedure of these analyses were part of our preregistration. We first subset the data to only include demonstration trials of participants. Next, we aggregated the data to the appropriate level for each measure of interest: trial-level for reward, move-level for behavioural congruency, and participant-level for the written strategies. 

In all models, the predictor of interest was a fixed effect of \textit{condition}, assessed by computing the bootstrapped confidence intervals around the estimate at 95\% confidence level. All confidence intervals are two-sided at the 95\% level. For the models predicting the measure across generations 1-4, we included random intercepts for \textit{participant} and \textit{population}, along with a random slope for \textit{generation} within \textit{population}. For models predicting measures within individual generations, we incorporated the random intercepts for \textit{participant} and \textit{population}. Reward outcomes were modeled using a linear link function, whereas behavioural congruency and strategies were modeled using a logistic link function. As preregistered, we centered and scaled the variable \textit{generation} to enhance model convergence and to make the main effect of \textit{condition} more interpretable. Residual diagnostics for the linear mixed-effects models confirmed approximate normality and homoscedasticity of residuals.

\subsection{Power analysis}

We determined the goal of 15 trees per condition by simulating data via an agent-based model resembling a variant of the model described in the next section. In this model, rates of learning success and individual discovery rates were calibrated based on pilot data. We created synthetic data for 100 independent runs of the full experimental design in this way. In a power analysis on this simulated data, we resampled runs with 10 or 15 populations, each 1000 times. We then checked the distribution of parameters for the main preregistered models 1a and 1b, and based our final decision on 1b (since it has the smaller subsample of the two main predictions). We found our predicted effect, as assessed by a t-value greater than 2 for "condition", in 957 (96\%) of the samples involving 15 populations, suggesting sufficient power to assess all predicted effects.

\subsection{Agent-based simulation}

We developed a parsimonious agent-based model to capture key mechanisms of multi-generational cultural transmission. Each generation consists of sequential individual learning, reward-based teacher selection, social transmission, strategy integration, demonstration, and reward updating.

Simulations comprised $M=50$ replications, $G=20$ generations, and $N_{\text{gen}}=20$ agents per generation. Depending on the condition, $N_{\text{m}}=5$ agents were machines with systematically manipulated exploration capacities $K_{\text{m}}$ and optimality biases $q_{\text{m}}$.

During individual learning, human agents performed $k_{\text{h}}$ trials, allocating exploration between optimal and myopic strategies according to an exploration bias $q_{\text{h}}$. Optimal and myopic strategies were discovered with cumulative probabilities determined by per-trial rates $d_{\text{opt}}$ (manipulated) and $d_{\text{myo}} = 0.5$, reflecting the greater accessibility of the myopic solution. Agents adopted an optimal, myopic, or null strategy, with priority given to the optimal strategy when both were discovered. Successful agents produced $K_{\text{demo}}=10$ demonstrations of the adopted strategy. Rewards were computed as
\begin{equation}
R = n_{\text{opt}}R_{\text{opt}} + n_{\text{myo}}R_{\text{myo}} + \epsilon_{\mathrm{eff}},
\quad
\epsilon_{\mathrm{eff}} \sim \mathcal{N}\!\left(0,\epsilon^2 K_{\text{demo}}\right),
\end{equation}
with $R_{\text{opt}}=1$, $R_{\text{myo}}=0.5$, and $\epsilon=0.1$.

The initial generation relied exclusively on individual learning. For generations $g>0$, learners sampled candidate teachers from the previous generation and selected among them via softmax weighting over rewards with temperature $\tau$. Social transmission occurred with probability $\lambda$. Individual and social learning outcomes were integrated such that optimal strategies dominated whenever acquired. Provided that evolutionary selection operates through teacher choice, we adopt this simple prioritization of demonstrated behavior, assuming that agents aim to maximize individual performance during demonstration.

We examined homogeneous and mixed populations across ranges of exploration bias, discoverability ($-\log_{10} d_{\text{opt}}$), and learnability ($1-\lambda$), and systematically varied machine capacity and bias. Additional analyses varied selection strength $\tau$, reward noise $\epsilon$, machine placement, population size, and machine prevalence to assess robustness. Full implementation details and extended analyses are provided in the Supplementary Information.

\section{Data Availability}
All experimental data are deposited on Zenodo (DOI: \href{https://doi.org/10.5281/zenodo.18547122}{10.5281/zenodo.18547122}) \cite{noauthor_code_2024}. The deposit contains both raw experimental data, as downloaded from the experiment interface, and processed data with derived variables used in the analyses. All data are de-identified and provided at the individual-participant and move level.

\section{Code Availability}
All code for reproducing the statistical analyses, visualisations, training of the machine player, and execution of the experiment is available on Zenodo (DOI: \href{https://doi.org/10.5281/zenodo.18547122}{10.5281/zenodo.18547122}) \cite{noauthor_code_2024}, including a static version of the experimental interface \cite{noauthor_experimental_nodate}.


\clearpage 

\bibliography{bibtex_2}
\bibliographystyle{unsrtnat}


\section{Acknowledgments}
We thank Jaeeun Shin and Sepideh Khalafi for their support with data collection. We are grateful to Ivan Soraperra for providing helpful feedback on the statistical method. 
Finally, we thank Peter Dayan and Wolfgang Hönig for their valuable feedback.

\section{Funding}
V.C. was funded by the Deutsche Forschungsgemeinschaft (DFG, German Research Foundation) under Germany's Excellence Strategy -- EXC 2002/2 ``Science of Intelligence'' -- project number 390523135. M.D. acknowledges funding from the French National Research Agency (Grant No. ANR-17-EURE-0010 and Grant No. ANR-21-CE28-0019-01).

\section{Author Contributions}
L.B., T.F.E., M.D., and I.R. conceived the research project and experimental methodology; L.B., T.F.E., A.-M.N., S.B., and V.C. refined the methodology; L.B., T.F.E., A.-M.N., M.D., and I.R. developed the theoretical grounding; T.F.E. pre-registered and conducted the formal analysis; L.B. conducted the agent-based simulation; S.B. developed the machine player; V.C. implemented the experiment; L.B., T.F.E., and A.-M.N. administered and supervised the study; L.B.,T.F.E., and A.-M.N. created the visualizations; L.B., T.F.E., A.-M.N., M.D., and I.R. wrote the paper.

\section{Competing Interests}
The authors declare no competing interests.


\newpage


\renewcommand{\thefigure}{S\arabic{figure}}
\renewcommand{\thetable}{S\arabic{table}}
\renewcommand{\theequation}{S\arabic{equation}}
\renewcommand{\thepage}{S\arabic{page}}
\setcounter{figure}{0}
\setcounter{table}{0}
\setcounter{equation}{0}
\setcounter{page}{1} 


\begin{center}
\section*{Supplementary Information for\\ \papertitle}

Levin~Brinkmann$^{\ast\dagger}$,
Thomas~F.~Eisenmann$^{\ast\dagger}$,
Anne-Marie~Nussberger$^{\dagger}$,\\
Maxime~Derex,
Sara~Bonati,
Valerii~Chirkov,
Iyad~Rahwan$^{\ast}$\\
\small$^\ast$Corresponding authors. Email: brinkmann@mpib-berlin.mpg.de (L.B.); eisenmann@mpib-berlin.mpg.de (T.F.E.); rahwan@mpib-berlin.mpg.de (I.R.)\\
\small$^\dagger$These authors contributed equally to this work.
\end{center}


\newpage


\section{Supplementary Text}
\section{Supplementary Note 1: Estimating Human Exploration Bias}
To estimate bias in human exploration of the networks, we constructed a matched random counterfactual for each human trajectory. We focused on two discrete
outcomes: the number of losses incurred along a solution path and the maximum
level reached in the network. For each category $k$ of an outcome variable $x$, we
quantified bias using a one-vs-rest log-odds ratio comparing the probability that a
human trajectory exhibits $x=k$ with the corresponding probability under the
random counterfactual:
\begin{equation}
\mathrm{LOR}(k)
= \log\!\left(\frac{P(x = k \mid \text{human})}{1 - P(x = k \mid \text{human})}\right)
- \log\!\left(\frac{P(x = k \mid \text{random})}{1 - P(x = k \mid \text{random})}\right).
\end{equation}
Thus, $\mathrm{LOR}(k)=0$ indicates no deviation from the random baseline, whereas
positive and negative values indicate over- and under-representation of category
$k$, respectively.

Uncertainty was assessed via nonparametric bootstrap resampling clustered at the
participant level, resampling all trajectories from the same participant
together. To avoid infinite log-odds in sparse categories, probabilities in each
bootstrap replicate were estimated using Jeffreys smoothing,
$\tilde p_k = (c_k + 1/2)/(n + 1)$, where $c_k$ is the number of trajectories with
$x=k$ out of $n$ in the corresponding condition. Confidence
intervals were obtained from the percentile range of the bootstrap distribution
of $\mathrm{LOR}(k)$.

\begin{figure}
    \centering
    \begin{subfigure}[t]{0.48\textwidth}
        \centering
        \includegraphics[width=\textwidth]{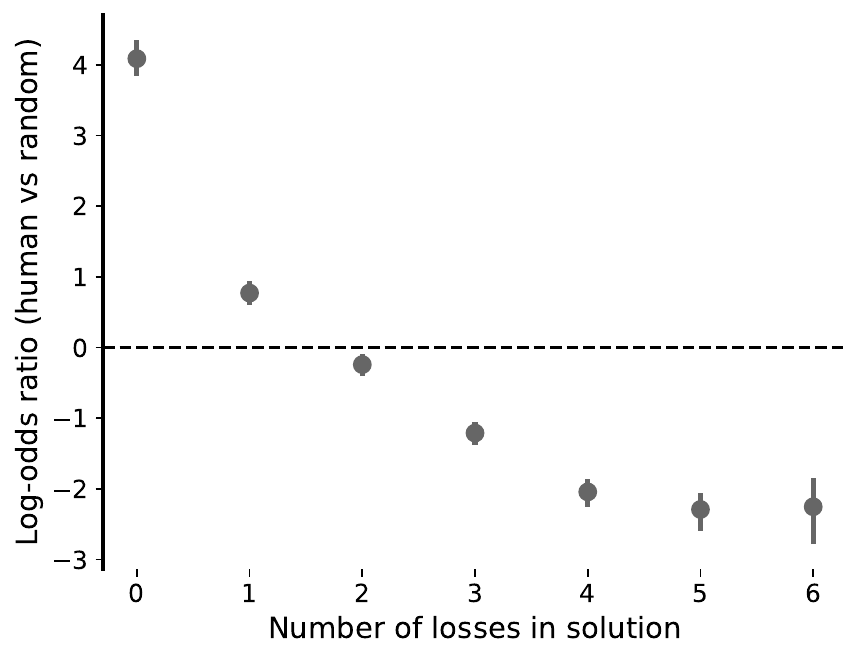}
        \caption{Number of losses.}
        \label{fig:lor_n_loss}
    \end{subfigure}\hfill
    \begin{subfigure}[t]{0.48\textwidth}
        \centering
        \includegraphics[width=\textwidth]{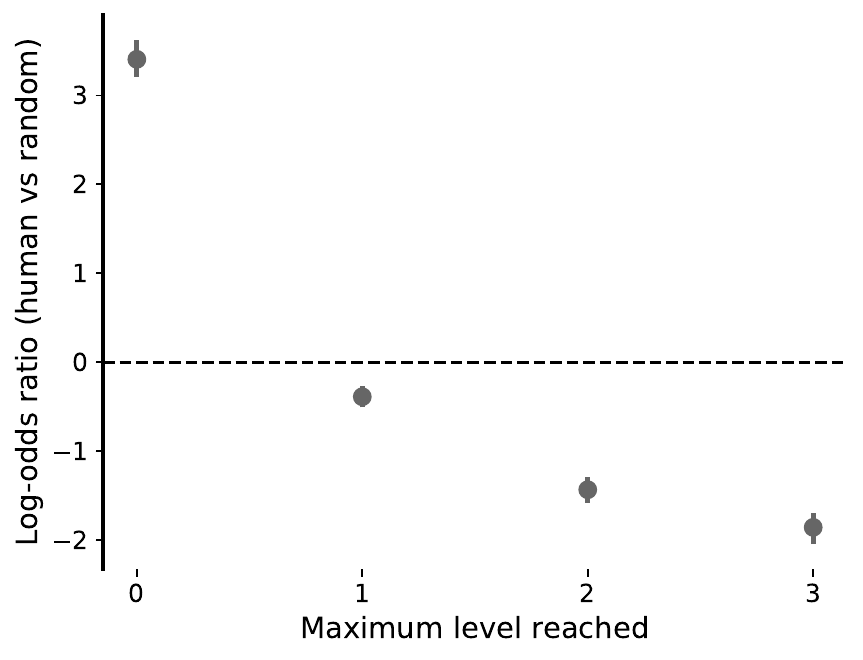}
        \caption{Maximum level reached.}
        \label{fig:lor_max_level}
    \end{subfigure}
    \caption{\textbf{Loss Aversion in Human Exploration.}
    Log-odds ratios comparing human and random-counterfactual distributions for
    (a) the number of losses incurred and (b) the maximum level reached, based on
    $N = 1{,}155$ participants. Points show $\mathrm{LOR}(k)$ and error bars
    indicate 95\% bootstrap confidence intervals (CI; clustered by participant and
    using Jeffreys smoothing). The dashed line at zero indicates no bias
    (i.e., the human distribution matches the random counterfactual).}
    \label{fig:lor_main}
\end{figure}

As shown in Fig.~\ref{fig:lor_n_loss}, the distribution of losses is highly
asymmetric relative to the random baseline. Zero-loss solutions are
over-represented ($\mathrm{LOR}(0)=4.39$, 95\% CI $[4.13,4.71]$; 1785 human vs.\ 51
random), corresponding to an increase in odds of more than two orders of
magnitude. In contrast, trajectories involving multiple losses are systematically
under-represented, with three losses
($\mathrm{LOR}(3)=-1.22$, 95\% CI $[-1.38,-1.06]$; 260 human vs.\ 736 random),
showing an approximately threefold reduction in odds.

Consistent with this pattern, Fig.~\ref{fig:lor_max_level} shows that human
trajectories are biased toward shallow solutions. Level~0 is
over-represented ($\mathrm{LOR}(0)=3.33$, 95\% CI $[3.16,3.53]$; 1785 human vs.\ 144
random), whereas higher levels are progressively under-represented, most notably
level~3 ($\mathrm{LOR}(3)=-1.86$, 95\% CI $[-2.03,-1.70]$; 192 human vs.\ 924
random). Together, these results indicate that aversion to incurring losses
constrains exploration to regions of the network with lower level.

\begin{figure}
    \centering
    \begin{subfigure}[t]{0.48\textwidth}
        \centering
        \includegraphics[width=\textwidth]{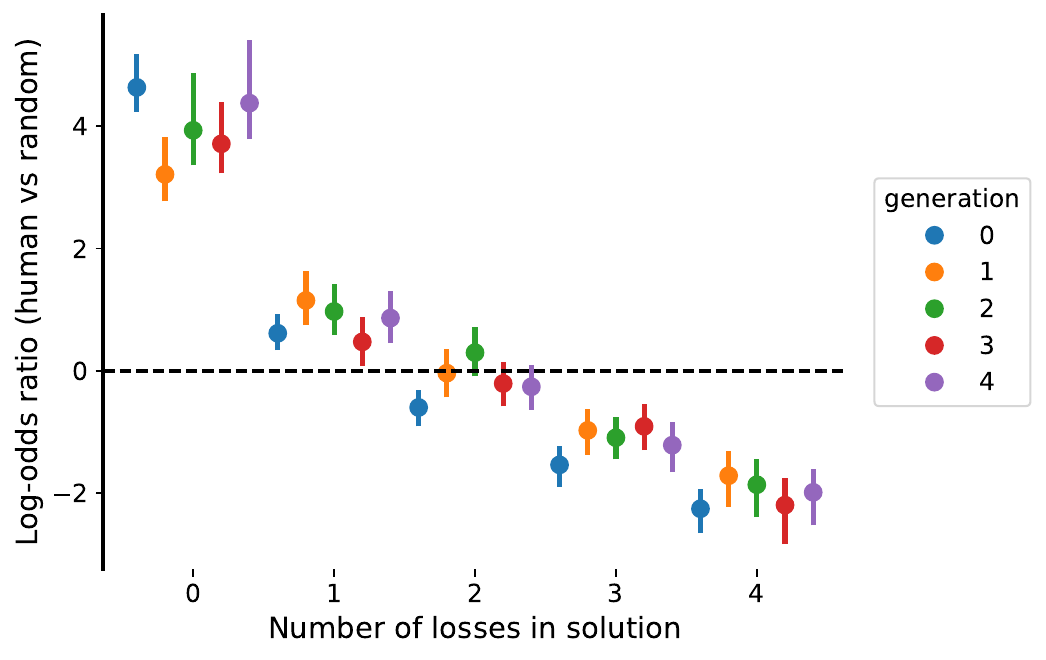}
        \caption{By generation.}
        \label{fig:lor_n_loss_gen}
    \end{subfigure}\hfill
    \begin{subfigure}[t]{0.48\textwidth}
        \centering
        \includegraphics[width=\textwidth]{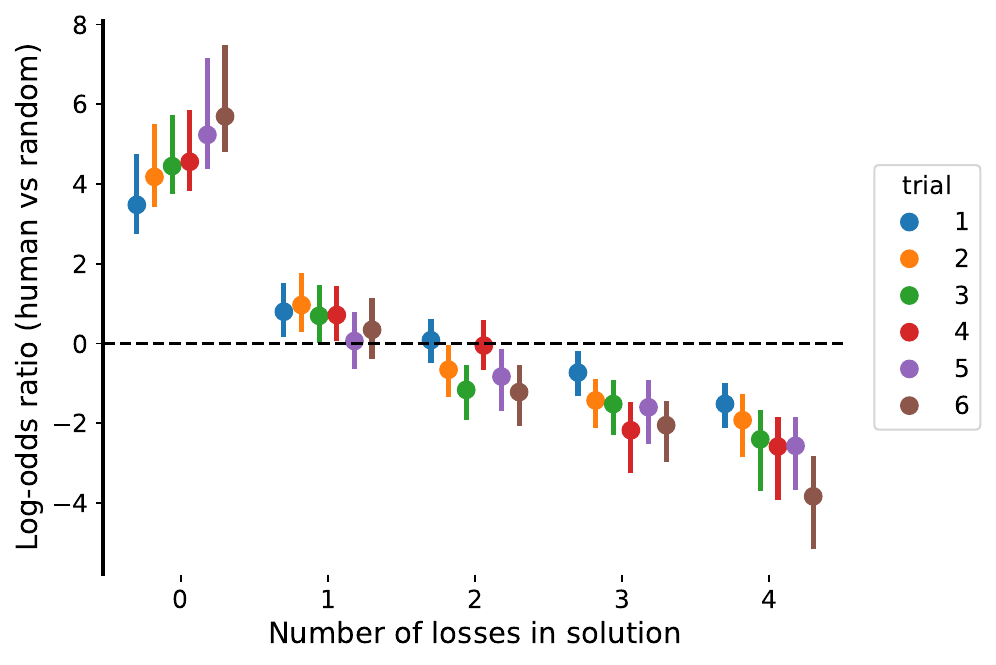}
        \caption{Across trials in generation~0.}
        \label{fig:lor_n_loss_trial}
    \end{subfigure}
    \caption{\textbf{Loss Aversion Across Generations and Trials.}
    Log-odds ratios for the number of losses, shown (a) separately for each
    generation ($n = 195$ human participants in generation~0 and $n = 240$ in each
    of generations~1--4) and (b) for each of the six learning trials in
    generation~0. Error bars indicate 95\% bootstrap CI (clustered by participant).
    The dashed line at zero indicates no bias (i.e., the human distribution matches
    the random counterfactual).}
    \label{fig:lor_dynamics}
\end{figure}

\begin{figure}
    \centering
    \includegraphics[width=1\textwidth]{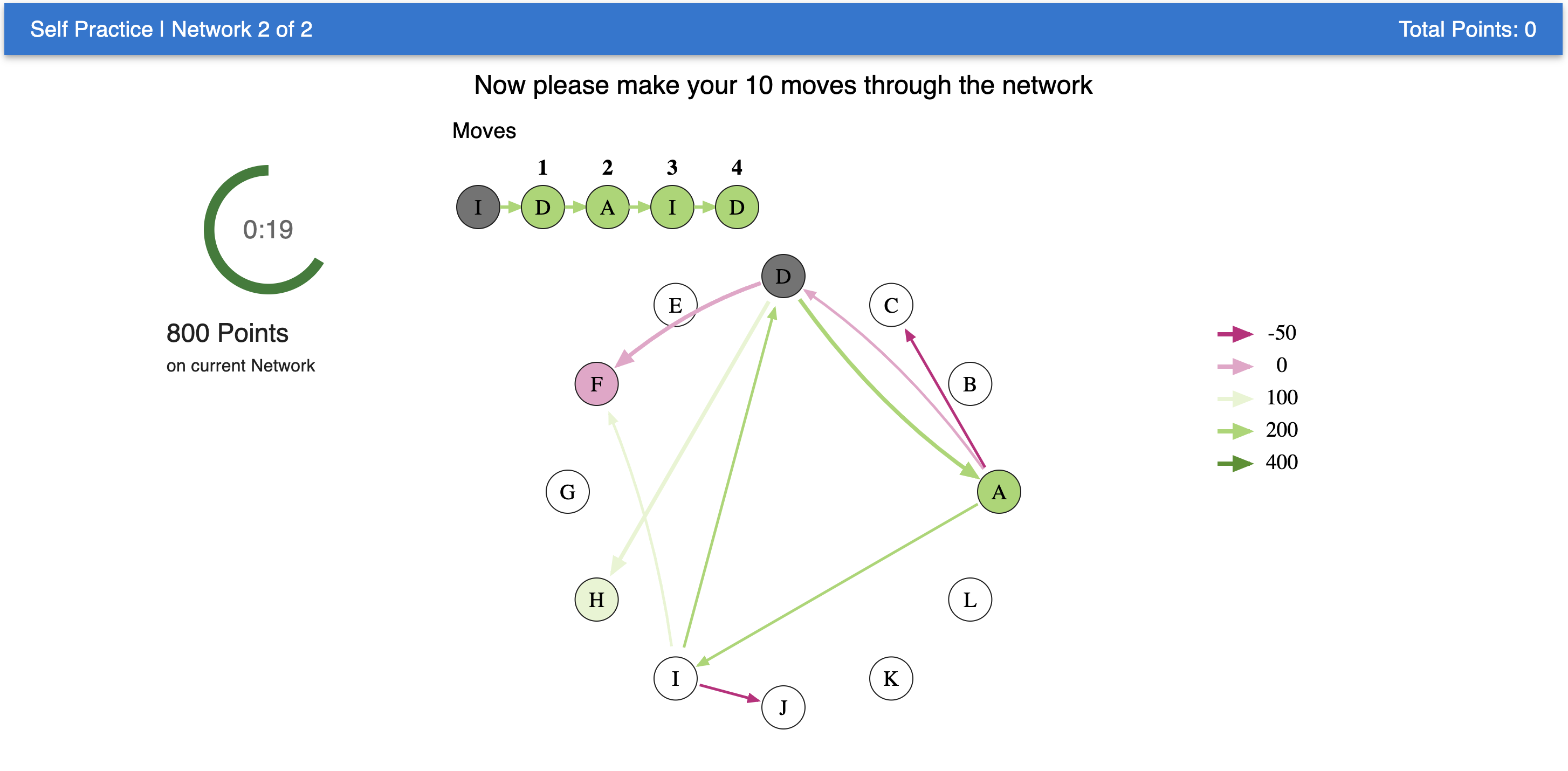}
    \caption{\textbf{Self-learning page.} In the first part of the experiment, participants are on their own to solve the full version of the task. This phase is of different length depending on the generation. In generation 1-4, participant can try themselves on two different networks. The main purpose here is for participants to get an better understanding of the task, before entering the social learning phase. In generation 0, participants have no social learning phase. Correspondingly, for these participants the self-learning phase is extended to a total of 6 networks. }
    \label{fig:self_practise}
\end{figure}

\begin{figure}
    \centering
    \includegraphics[width=1\textwidth]{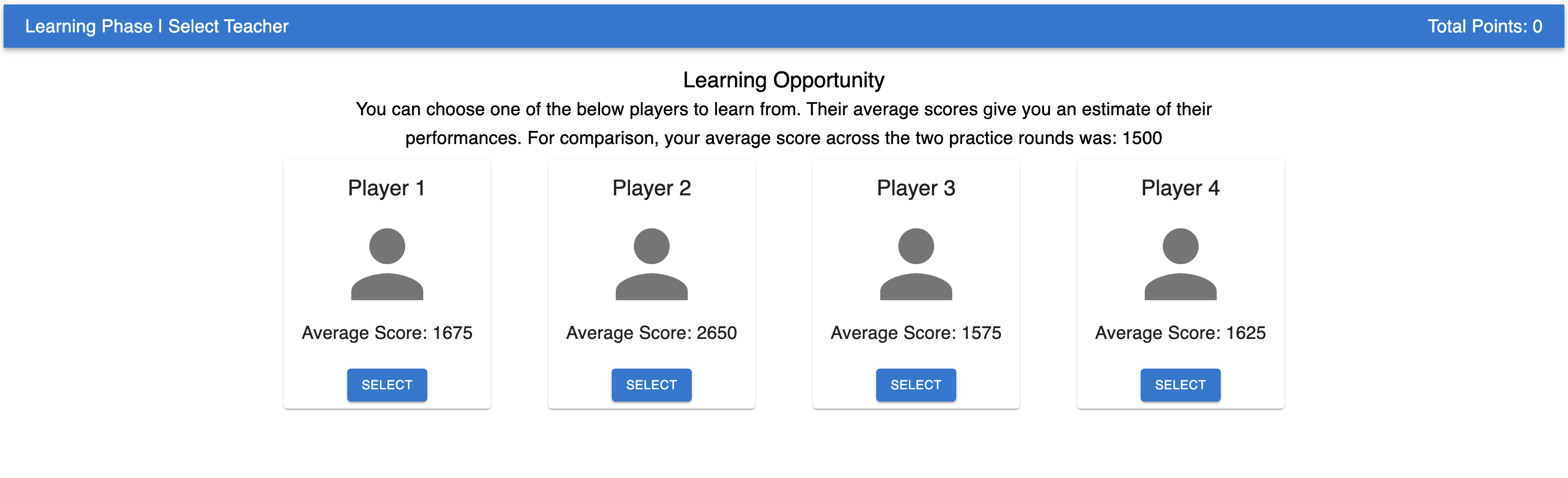}
    \caption{\textbf{Select demonstrator.} Before entering the social learning phase, participants can select one out of five potential demonstrators. The only information guiding participants' selection is the demonstrators' average score, which is computed from the points collected during the demonstrators' demonstration phase. Otherwise, each demonstrator is symbolised identically, with no discrimination between human and machine player. Participants were informed that demonstrators could be humans or machines, however. As an additional comparison to make their choice, participants' own average performance on the previous two self-learning networks is displayed. }
    \label{fig:select_teacher}
\end{figure}

\begin{figure}
    \centering
    \includegraphics[width=1\textwidth]{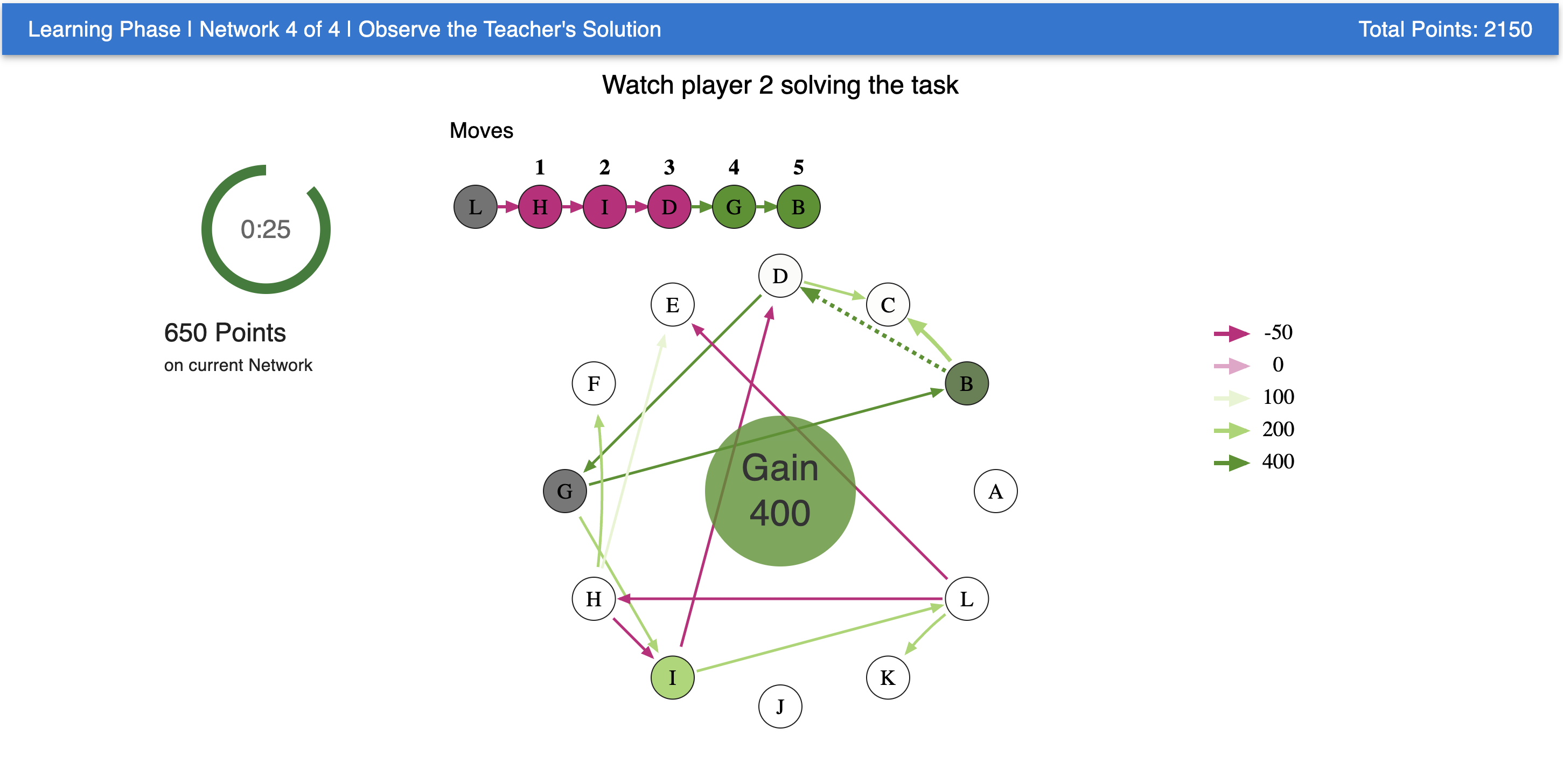}
    \caption{\textbf{Social learning by observation.} Participants see a replay of the demonstrator solving the task. The task is displayed in the same way as during self-play, however the UI is deactivated. For each step of the solution, the next move is indicated as a dashed line. }
    \label{fig:observation}
\end{figure}

\begin{figure}
    \centering
    \includegraphics[width=1\textwidth]{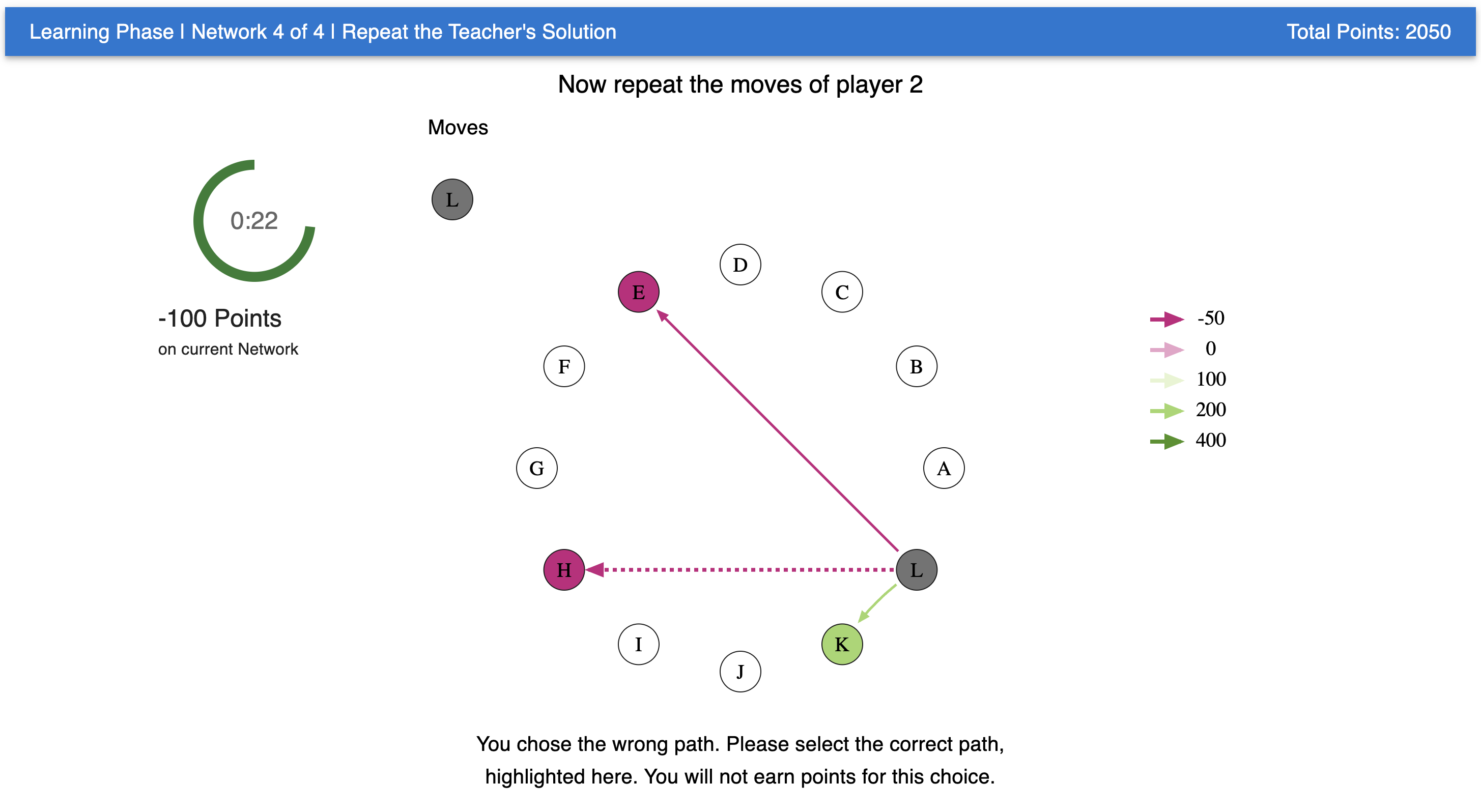}
    \caption{\textbf{Social learning by repetition.} Participants are exposed to the same network and are asked to repeat the demonstrators' moves. Participant are receiving +100 points for correct repetitions and -100 points for wrong repetitions. If a wrong move is chosen, the right move is indicated by a dashed line. Participants still have to enact the right move, but do not receive points. }
    \label{fig:repeat}
\end{figure}

\begin{figure}
  \centering
  \includegraphics[width=1\textwidth]{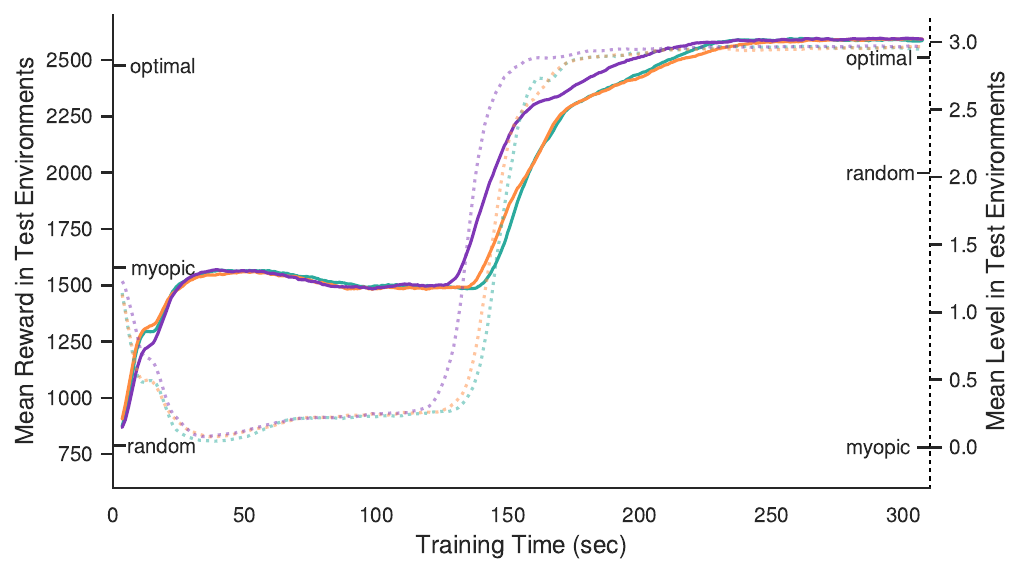}
\caption{\textbf{Algorithmic performance during training}. We trained three Deep Q-learning algorithms – signified by the three distinct colours – during a time period that was similar to the Individual Learning Phase for human participants (i.e., 315 seconds) for 5,000 episodes on 10,000 unique networks (distinct from the experimental ones). The three colors correspond to three independent training runs with different random seeds (seeds 0, 1, and 2). The left y-axis references the solid lines which denote the algorithmic agents' achieved rewards. The right y-axis and dashed lines denote the achieved maximum level, where the highest level could only be reached by incurring losses to progress. We marked the average values for three heuristic benchmarks on both y-axes: a random agent, a myopic agent always choosing the highest immediate reward, and a loss-seeking agent pursuing losses when possible. During training, the algorithms first discover the myopic strategy, before finally converging on the optimal strategy.}
\label{fig:algorithm}
\end{figure}

As shown in Fig.~\ref{fig:lor_n_loss_gen}, the qualitative pattern of loss aversion
is stable across generations: zero-loss solutions are consistently
over-represented, whereas trajectories involving multiple losses are
under-represented. Participants in generation 0 had six exploration trials, whereas participants in later generations had only two. We find no statistically significant evidence that the additional trials attenuated this bias.

Within generation~0 (Fig.~\ref{fig:lor_n_loss_trial}), the bias against incurring losses increases over successive trials. Zero-loss solutions become
progressively more over-represented, while trajectories with several losses are
increasingly suppressed, indicating growing loss aversion even in the absence of
inter-generational learning.

\section{Supplementary Note 2: Classification of Populations}

Figure \ref{fig:metrics_individual_level} displays the average points of all participant in each population and generation. In the reward network task, a point value above 2000 is only possible if the largest level 3 is reached and multiple largest rewards of 400 points are collected. Therefore, we classify a generation of a population as 'discovered' if the average point value of the top performer exceeds 2000 points. Focusing on the last generation of each population, we found 9 groups exceeding this threshold in the human-machine population and one group in the human population.

Correspondingly, we found:

\begin{itemize}
    \item \textbf{Permanently preserved:} In 9 out of 15 human-machine populations, humans permanently preserve the machine discovery.
    \item \textbf{Temporarily preserved:} In 6 out of 15 human-machine populations, humans temporarily preserve the machine discovery. This means that the machine discovery was preserved at least in the second generation; however, no participant in the last generation exceeded the threshold.
    \item \textbf{Discovered:} In 1 out of 15 human populations, the optimal strategy was discovered, and the maximum average point value in the last generation exceeds 2,000 points.
    \item \textbf{Not discovered:} In 14 out of 15 human populations, no participant exceeded an average point value of 2,000.
\end{itemize}

\section{Supplementary Note 3: Qualitative Summary of Prototypical Strategies}

Many written strategies referred to “looping” as a feature of the task that was related to higher performance, but not necessarily optimal: \squote{the trick is to identify a path between two nodes that both award high points and then repeat the path back and forth.}, \squote{Find two letters with connecting green arrows and go back and forth between them.} Some participants admitted that they did not have a good plan of what to do: \squote{no idea seriously just choosing patterns}, \squote{Unsure of an effective strategy}. Others confidently reported “superstitions” that focused on parts of the task that had no actual relation to good performance: \squote{remember the first random letters then remember a sequence}, \squote{Longer routes seemed to provide more options for success}.

Frequently, written strategies explicitly referred to myopic behavior: \squote{Ignoring the big loss moves}, \squote{Always avoiding neutral or negative paths seems to be the best strategy.} Sometimes, these myopic strategies even explicitly highlighted participants’ bias against early losses: \squote{always choose the line that gives you points instead of taking away points for less risk}, \squote{I just followed the green paths, I don't even really know what else I would try to do? Picking the negative point options makes no sense, after all.} In contrast, participants that had learned the loss strategy frequently referred to it explicitly and showed their long-term planning: \squote{Take a loss for the first three moves then move between two high scoring nodes.} \squote{if you take a hit early (eg the first 3 nodes are -50), you are in a position to maximize the following 7 nodes, each with +400. i tried it differently initially, but player 2's method is best.} The last example also emphasizes the important role of social learning for developing this participant’s strategy.

Last, the one exceptional human player in the human populations that discovered the loss strategy on their own wrote \squote{I try o track my best point} (sic) after training, maybe indicating an approach that was open to exploring while systematically assessing their point score. After social learning, they wrote \squote{I tried to keep losing till I see the thinker green line.} (sic), clearly indicating the loss strategy even though no previous participant was able to demonstrate it to them. Presumably, the participant made good use of the “try yourself” trials to explore also unusual options, and discovered the loss strategy this way, as the only player to do so in the 15 human populations.

One curious observation is that 7 out of 120 participants in generation 0 in human populations got coded as “mentioned the loss strategy” in the data. Writing down and coding written strategies is an inherently interpretative process, and it is unclear whether these participants just voiced suspicions, actually figured out the strategy, or whether this is just noise in the data or in the coding process. A qualitative inspection of the strategies suggests that they reasonably reflect the loss strategy, at least as defined by our coding manual. Most participants remained vague, e.g. “you have to reduce some points to gain some”, suggesting that at least the exact implementation of taking three losses might have been hidden from them. 
Most importantly, the hunch that they might have had did not translate into better performance. The 9 players range from 1000 to 1600 points total, whereas we expected more than 2000 points for a player implementing the loss strategy (see “Humans Preserve the Machine Discovery”). Crucially, also their demonstration trials do not look much better, with only two select trials just scratching the 2000 point mark, but many failed attempts demonstrating 1500 points or less. So even by the end of the experiment, they were not able to demonstrate that they learned a superior strategy.
\section{Supplementary Note 4: Task Performance (robustness checks)}

Figure \ref{fig:metrics_individual_level} shows the average reward per participants, separated by generation and population. Descriptively, human participants in the average human population started out equal to the average human-machine population in generation 0, but human-machine populations earned consistently higher rewards in generations 1 to 4. 

We ran the linear mixed-effects model testing prediction 1a ("General Benefit of AI") on the demonstration trials of generations 1 to 4, with \textit{total reward} as the dependent variable and fixed effects for \textit{condition}, \textit{generation}, and their interaction. Additionally, we included random intercepts for \textit{participant} and \textit{population}, as well as a random slope for \textit{generation} on \textit{population}. As preregistered, we centered and scaled \textit{generation} to improve model convergence and make the main effect for \textit{condition} interpretable. In line with our prediction, the model revealed a significant main effect for \textit{condition} ($\beta = -309.2$, \textit{SE} = $57.0$, $\text{CI}_{95\%}$ = [$-418.4$, $-200.1$]), indicating that human-machine populations outperformed human populations overall. 

The model for prediction 1b ("Long-lasting Benefit of AI") followed a similar logic, but was using only the data of generation 4. Given the more restricted data set, only the fixed effect for \textit{condition} and the random intercepts for \textit{participant} and \textit{population} remained. Similar to the first model, there was a meaningful main effect for \textit{condition} ($\beta = -229.1$, \textit{SE} = $82.4$, $\text{CI}_{95\%}$ = [$-383.1$, $-55.5$]), supporting our prediction that human-machine populations would perform better than human populations by the last generation.

The model for prediction 2a ("Transmission from AI to Humans") was identical to the one for 1b, but used the data of generation 1. Again, there was a significant main effect for \textit{condition} ($\beta = -404.6$, \textit{SE} = $59.4$, $\text{CI}_{95\%}$ = [$-524.8$, $-286.1$]), supporting our prediction that human-machine populations performed better than human populations in the immediate generation learning from machine agents. Here, the model returned a warning about a singular fit, so we also checked all other optimizers available in \texttt{lme4}; however, since all optimizers converged to identical values for the fixed effects, we concluded the warning to be false-positive.

\subsection{Human-Machine Behavioural Congruency}

Figure \ref{fig:metrics_individual_level}.B shows the proportion of human moves congruent with machine-generated moves per demonstration trial, summarized for each participant. Qualitatively, the results are similar to the results for rewards: no statistically significant difference between conditions in generation 0, but there is consistently more behavioural congruency in human-machine populations for generations 1 to 4. 

The models assessing predictions 1a to 2a were identical to the ones for reward, with exception of the different outcome measure and being on the level of moves per trial. Since behavioural congruency was a binary variable, we used a logit link here. All three models replicated the expected effects for condition found for reward on the behavioural congruency measure: Human-machine populations were behaviourally more congruent than human populations overall ($\beta = -0.87$, \textit{SE} = $0.14$, $\text{CI}_{95\%}$ = [$-1.12$, $-0.61$]), in the last generation ($\beta = -0.64$, \textit{SE} = $0.20$, $\text{CI}_{95\%}$ = [$-1.02$, $-0.25$]), and in the first generation ($\beta = -1.13$, \textit{SE} = $0.13$, $\text{CI}_{95\%}$ = [$-1.41$, $-0.88$]). Again, the last model returned a warning about a singular fit, but all available optimizers converged to identical fixed effects until the second digit, so we took it as a false positive.

\subsection{Written Strategies}

Figure \ref{figure3}.C  shows the proportion of human participants reporting the loss strategy, summarized by generation and population. In the human populations, we observe practically no mention of the loss strategy, with proportions close to 0 throughout all 5 generations; the exception to this is the single population that discovered the loss strategy towards the end, resulting in 50\% of the participants referring to the strategy in generation 4. In contrast, participants in the human-machine populations consistently mentioned the loss strategy explicitly, with proportions of about 30\% to 50\% per generation. In generations 1 to 4 (i.e. with social learning), almost no participants mentioned the loss strategy immediately after the training trials (human populations: 1.7\%; human-machine populations: 2.7\%), but 39.8\% of participants in the human-machine populations mentioned it after social learning had occurred (compared to 1.9\% in the human populations). In comparison, in generation 0 (without social learning) very few participants in either condition referred to the loss strategy at either time point (human-machine t1: 1.3\%; t2: 1.3\%; human-only t1: 2.5\%; t2: 5.8\%).

The model assessing prediction 2b ("Explicit Recognition of Counterintuitive Strategy") was identical to the one for prediction 1a, but using a logit link due to the binary outcome variable and missing the random intercept for participants, since it was built on participant level. Again, there was a significant main effect for condition ($\beta = -4.18$, \textit{SE} = $0.62$, $\text{CI}_{95\%}$ = [$-6.30$, $-3.11$]), indicating that human-machine populations explicitly mentioned the loss strategy more frequently than human populations from generations 1 to 4. This model also returned a warning about a singular fit, but all available optimizers converged to qualitatively similar parameters.

\subsection{Robustness Checks}

As a robustness check, we revisited the random-effects structure in Models 1a and 2b. In the original specification, generation was treated as a continuous predictor and incorporated via a random slope, but not explicitly modeled as a grouping factor. We therefore re-estimated the models with generations nested within populations, replacing the fixed effect of generation. In Model 1a, participants were additionally nested within generations (this structure was not feasible in Model 2b). All models converged under this alternative specification, and the main effect of condition remained robust:

\begin{itemize}
  \item \textbf{Performance Model 1a}: $\beta = -309.2$, \textit{SE} = $40.3$, $\text{CI}_{95\%}$ = [$-418.0$, $-195.4$]

\item \textbf{Congruency Model 1a}: $\beta = -0.87$, \textit{SE} = $0.14$, $\text{CI}_{95\%}$ = [$-1.14$, $-0.60$]

\item \textbf{Written Strategy Model 2b:} $\beta = -4.24$, \textit{SE} = $0.59$, $\text{CI}_{95\%}$ = [$-5.91$, $-3.29$]

\end{itemize}
\section{Supplementary Note 5: Agent-Based Model}

We developed a parsimonious agent-based model capturing the essential components of our multi-generational cultural transmission experiment. Each generation implements the following sequential processes: individual learning, reward-based teacher selection, social transmission, strategy integration, demonstration, and reward updating.

Each simulation consists of $M = 50$ independent replications, $G = 20$ generations, and $N_{\text{gen}} = 20$ agents per generation. Depending on the experimental condition, a subset $N_{\text{m}} = 5$ of agents are labeled as machines, characterized by distinct exploration budgets $K_{\text{m}}$ and optimality biases $q_{\text{m}}$, which are systematically manipulated.

\paragraph{Individual learning.}
Each agent labeled as human performs $k_{\text{h}} = 10$ learning trials, allocated between exploration of optimal and myopic strategies according to an exploration bias $q_{\text{h}} = 0.01$, chosen to investigate a regime with a considerable bias toward the myopic solution:

\begin{equation}
k_{\text{opt}} \sim \text{Binomial}(k, q_{\text{opt}}), \quad
k_{\text{myo}} = k - k_{\text{opt}} .
\end{equation}

Let $d_{\text{opt}}$ denote the per-trial discovery rate of the optimal strategy, which is experimentally manipulated, and let $d_{\text{myo}} = 0.5$ denote the discovery rate of the myopic strategy. The cumulative discovery probabilities are given by:

\begin{equation}
D_{\text{opt}} = 1 - (1 - d_{\text{opt}})^{k_{\text{opt}}}, \quad
D_{\text{myo}} = 1 - (1 - d_{\text{myo}})^{k_{\text{myo}}}.
\end{equation}

Agents stochastically discover either an optimal solution, a myopic solution, or no solution. If both solutions are discovered, priority is given to the optimal solution.

\paragraph{Demonstration and rewards.}
Agents produce $K_{\text{demo}}$ independent demonstrations. Let $n_{\text{opt}}$ and $n_{\text{myo}}$ be the numbers of optimal and myopic demonstrations (the remainder are null). Each demonstration is perturbed by independent noise $\epsilon_k \sim \mathcal{N}(0,\epsilon^2)$. The total reward is
\begin{equation}
R = n_{\text{opt}} R_{\text{opt}} + n_{\text{myo}} R_{\text{myo}} + \epsilon_{\text{eff}},
\quad
\epsilon_{\text{eff}}  \sim \mathcal{N}\!\left(0,\epsilon^2 K_{\text{demo}}\right),
\end{equation}
where $R_{\text{opt}}=1$, $R_{\text{myo}}=0.5$, and $\epsilon=0.1$.

\paragraph{Teacher selection and social learning.}
For generations $g > 0$, each learner samples $k$ candidate teachers uniformly at random from the previous generation. Teachers are selected according to a softmax function over their observed rewards:

\begin{equation}
p_i \propto \exp(r_i / \tau),
\end{equation}

where $\tau$ controls the intensity of selection.

Learners adopt the selected teacher’s optimality state with probability $\lambda$. Individual and social learning outcomes are integrated according to:

\begin{equation}
s_g = \max(s_{\text{ind}}, s_{\text{social}}),
\end{equation}

ensuring that optimal strategies dominate myopic or null strategies whenever they are acquired. In the initial generation ($g = 0$), learning proceeds exclusively through individual exploration.

\paragraph{Simulated conditions.}

We first investigated the ability of homogeneous populations to discover and maintain the optimal strategy as a function of exploration capacity ($k_{\text{h}}$), bias toward myopic exploration ($q_{\text{h}}$), optimal solution discoverability ($-\log_{10} d_{\text{opt}}$), and social learnability ($1 - \lambda$). In Figure~\ref{fig.bias_discoverability}, we report the probability of discovery across populations persisting for 1, 5, and 20 generations.

We then focused on scenarios with heterogeneous exploration capacity and bias, in which the majority of agents (i.e., human agents) are limited in their exploration capacity ($k_{\text{m}} = 10$) and exhibit a bias ($q_{\text{m}} = 0.01$) toward myopic solutions. We introduced five machine agents in the first generation and systematically manipulated their exploration capacities ($k_{\text{m}} \in {10, 1000, 100000}$) and biases toward optimal strategies ($q_{\text{m}} \in {0.01, 0.5}$). Figure~\ref{fig.abm} shows the fraction of agents adopting the optimal strategy in the final generation as a function of social learnability ($1 - \lambda$) and optimal solution discoverability ($-\log_{10} d_{\text{opt}}$).

Next, we varied the strength of social learning selection $\tau \in [0.1, 0.2, 0.4]$ and examined different machine-placement scenarios (no machines, machines only in the first generation, and machines in all generations; see Figure~\ref{fig.abm_social_learning_selection}). Similarly, we varied the noise level applied to individual realized rewards $\epsilon \in [0.1, 0.2, 0.4]$  (see Figure~\ref{fig.abm_reward_uncertainty}).

Finally, we assessed the robustness of the main findings for high-capacity $k_{\text{m}} = 100000$, unbiased $q_{\text{m}} = 0.5$ machines  with respect to the number of machines in the first generation $N_{\text{m}} \in [1, 5]$ and the total population size per generation $N_{\text{total}} \in [20, 100]$ (see Figure~\ref{fig.abm_robustness}). 

\section{Supplementary Note 6: Pilot Studies}

We run 10 cycles of pilots. Each in each cycle we run some of a set four different types of pilots indicated by the letters A to D. These type of pilots differ in the experimental setup and therefore allowed to test different premises.

\begin{itemize}
    \item \textbf{Pilot type A}: Participants are placed in generation 0 of the experiment. They therefore have no chance of social learning. This pilot allows conclusion about the discoverablity of the loss strategy.
    \item \textbf{Pilot type B}: Participants are placed in the first generation  of the experiment. Solutions of the first generation are provided by rule based algorithm. Generally, we provided participants with a mix of teachers. Some of the teacher would show a solution of the loss strategy while other teacher would display a myopic solution. This pilots allows to infer about the learnability of the solution, provided the participant is exposed to the loss strategy. This type of pilots also allows to infer about the selection of a teacher by the participant. 
    \item \textbf{Pilot type C}: This pilot type was similar to pilot type B, however, participant could only choose from myopic solutions to learn from. 
    \item \textbf{Pilot type D}: This pilot type was the most complex of all. Here we collected data from the 0th and 1st generation. No machine solution was provided. Like pilot A, this pilot was intended to investigate the learnability. Additionally this pilot allowed a technical check of the multi-generational assignment. 
\end{itemize}

Not all pilots where informative. In the following we present a selection of all pilots.

\begin{itemize}
    \item \textbf{Pilot 1 A / B}
    \begin{itemize}
        \item \textbf{Setup}: The network task presented in these earlier pilots is similar to the final reward network task, with important differences:
        \begin{itemize}
            \item The full network is shown to participants, but humans are expected to behave suboptimally due to cognitive constraints in exploring the full solution space.
            \item The network is smaller, with only 8 nodes.
        \end{itemize}
        \item \textbf{Results}: Participants fail to discover the loss strategy. Even when shown the loss strategy, they fail to repeat it.
    \end{itemize}
    \item \textbf{Pilot 2 A / B}
    \begin{itemize}
        \item \textbf{Setup}:
        \begin{itemize}
            \item Increased the time for participants.
            \item Made it easier for participants to reproduce the correct strategy by showing it in written form.
            \item Asked participants to write down their strategies.
        \end{itemize}
        \item \textbf{Results}:
        \begin{itemize}
            \item In Pilot 2 A, participants enacted and described a myopic strategy.
            \item In Pilot 2 B, only 4 out of 10 participants were able to reproduce the loss strategy, either in practice or verbally, even after explicit instruction.
        \end{itemize}
    \end{itemize}
    \item \textbf{Pilot 3 B}
    \begin{itemize}
        \item \textbf{Setup}: Focused on the learnability of the loss strategy:
        \begin{itemize}
            \item Intensified the social learning phase.
            \item Introduced iterative observation and try-yourself pages, allowing participants to play the task themselves and compare their performance to the teacher’s.
        \end{itemize}
        \item \textbf{Results}: Only 2 out of 10 participants somewhat learned the loss strategy.
    \end{itemize}
    \item \textbf{Pilot 4 B}
    \begin{itemize}
        \item \textbf{Setup}: Made the written correct strategy even more salient.
        \item \textbf{Results}: No significant improvement in learnability.
    \end{itemize}
    \item \textbf{Pilot 5 B}
    \begin{itemize}
        \item \textbf{Setup}:
        \begin{itemize}
            \item Simplified the task.
            \item Redesigned the networks to better separate the myopic and loss strategies in terms of point values.
        \end{itemize}
        \item \textbf{Results}: Most participants learned the loss strategy and were able to reproduce it in demonstration trials.
    \end{itemize}
    \item \textbf{Pilot 5 A / C}
    \begin{itemize}
        \item \textbf{Setup}:
        \begin{itemize}
            \item Conducted two more pilots with the same experimental design as Pilot 5 B.
            \item In Pilot 5 C, participants could socially learn but only had myopic solutions to choose from.
            \item In Pilot 5 A, participants learned only individually.
        \end{itemize}
        \item \textbf{Results}: Performance in both pilots was mixed. While fewer participants used the loss strategy, some did. This was deemed a failure, as the goal was for no (or very few) participants to discover the loss strategy on their own.
    \end{itemize}
    \item \textbf{Pilot 6 B}
    \begin{itemize}
        \item \textbf{Setup}:
        \begin{itemize}
            \item Redesigned the task and experimental flow to prevent backward induction.
            \item Made unexplored parts of the networks hidden, reflecting real-world constraints like those in Go.
            \item Enhanced the social learning phase to highlight the exploration advantage of the machine.
        \end{itemize}
        \item \textbf{Results}: Promising results, with 3 out of 10 participants understanding the loss strategy but struggling to execute it.
    \end{itemize}
    \item \textbf{Pilot 7 B}
    \begin{itemize}
        \item \textbf{Setup}: Improved the UI in the social learning phase to:
        \begin{itemize}
            \item Make potential strategies more salient when the teacher shows the solution.
            \item Emphasize the contrast between participants’ suboptimal solutions and the teacher’s optimal solution.
        \end{itemize}
        \item \textbf{Results}: Five out of 10 participants somewhat learned the strategy, but many struggled to reproduce it.
    \end{itemize}
    \item \textbf{Pilot 8 A / B}
    \begin{itemize}
        \item \textbf{Setup}: Redesigned the network task to:
        \begin{itemize}
            \item Remove links from higher levels back to the first level, which tempted participants to avoid losses.
            \item Ensure losses only occurred when moving up levels, with the largest gains exclusive to the highest level.
        \end{itemize}
        \item \textbf{Results}:
        \begin{itemize}
            \item Pilot 8 B: Six out of 10 participants successfully implemented the loss strategy, with three doing so without error.
            \item Pilot 8 A: Confirmed that participants could not discover the loss strategy on their own.
        \end{itemize}
    \end{itemize}
    \item \textbf{Pilot 9 D}
    \begin{itemize}
        \item \textbf{Setup}: Tested the task with more participants (N = 51) to ensure the strategy was not discoverable.
        \item \textbf{Results}: None of the participants discovered the loss strategy.
    \end{itemize}
    \item \textbf{Pilot 9 B}
    \begin{itemize}
        \item \textbf{Setup}: Ran another 10 participants in the B configuration.
        \item \textbf{Results}: Only 7 out of 20 participants learned the strategy. Some networks where the loss strategy was not superior to the myopic strategy confused participants.
    \end{itemize}
    \item \textbf{Pilot 10 B}
    \begin{itemize}
        \item \textbf{Setup}: Removed networks where the loss strategy performed worse than the myopic strategy.
        \item \textbf{Results}: Nine out of 20 participants imitated the loss strategy without error.
    \end{itemize}
\end{itemize}

Through this piloting, we calibrated the task such that the optimal strategy's complexity was not trivial but also not impossible for human minds to discover and feasible to learn. For instance, in an early set of pilot studies, a few participants were able to discover the optimal strategy through backwards induction. To more realistically reflect that backwards induction is prohibitively costly and/or impossible due to the sequential ordering of effects in real-world problems, our final task consisted of networks in which paths were revealed sequentially according to players' moves. Illustrating the defining influence of learnability, in pilots where the optimal strategy involved incurring a precise number of large losses (e.g., no more or less than three), participants struggled to learn this within the limited time available in our experiment. Our final task thus involved a simplified optimal strategy of taking losses whenever presented with one. 
\newpage

\begin{figure}
  \centering
  \includegraphics[width=0.6\linewidth]{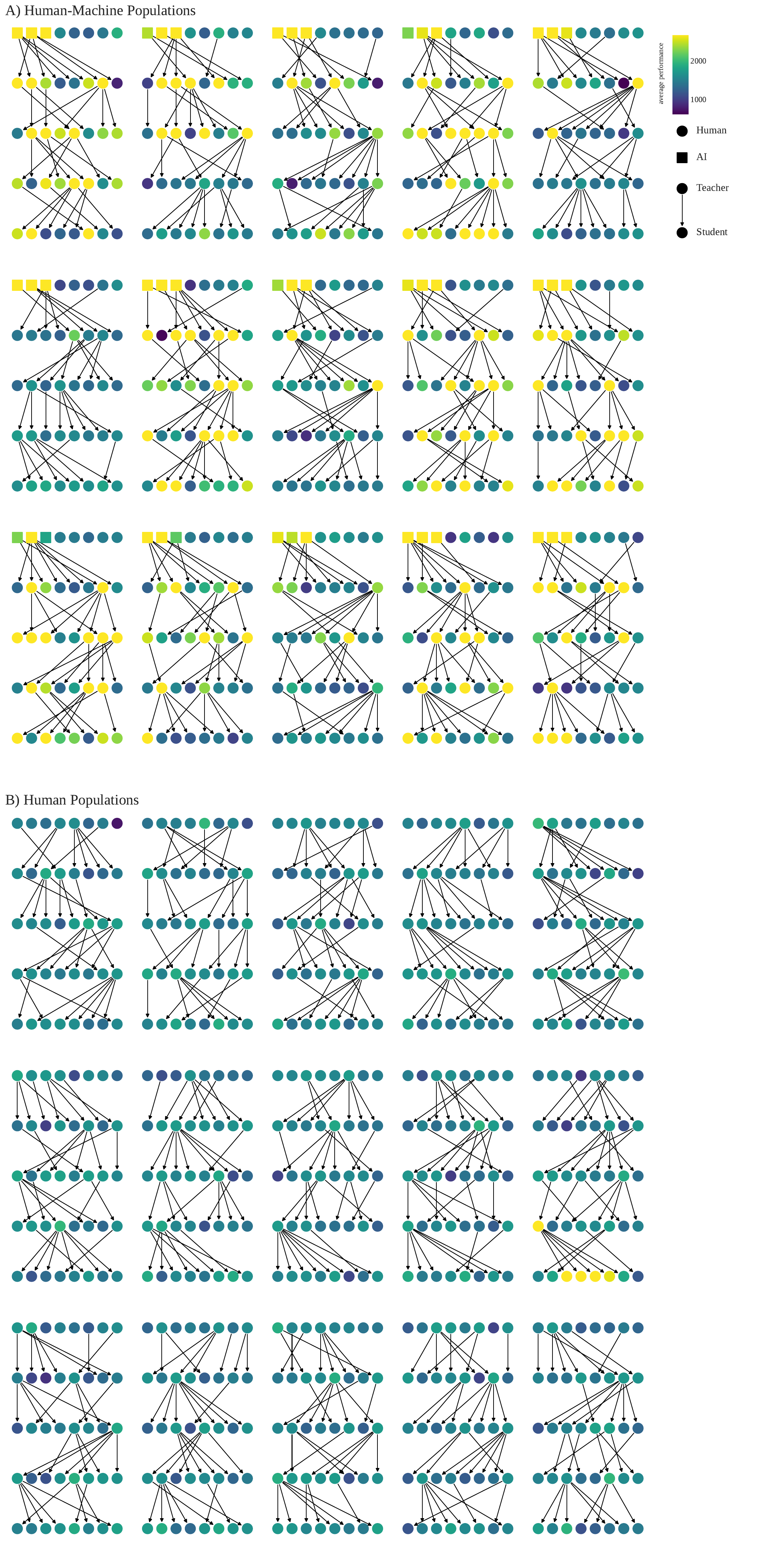}
\caption{\textbf{Performance of All 30 Populations}: Each population consists of five generations, each with eight players. In human-machine populations, there are three machine players in the first generation. Human players are represented by circles and machine players by squares, with the color of each symbol indicating the player's average score. The color scale spans the empirical range of average scores across all players, with lighter shades corresponding to lower scores and darker shades to higher scores. In generations following the first, each player selects a single teacher. Arrows depict these teacher-student relationships.}

  \label{fig:populations_score}
\end{figure}

\begin{figure}
  \centering
  \includegraphics[width=0.6\linewidth]{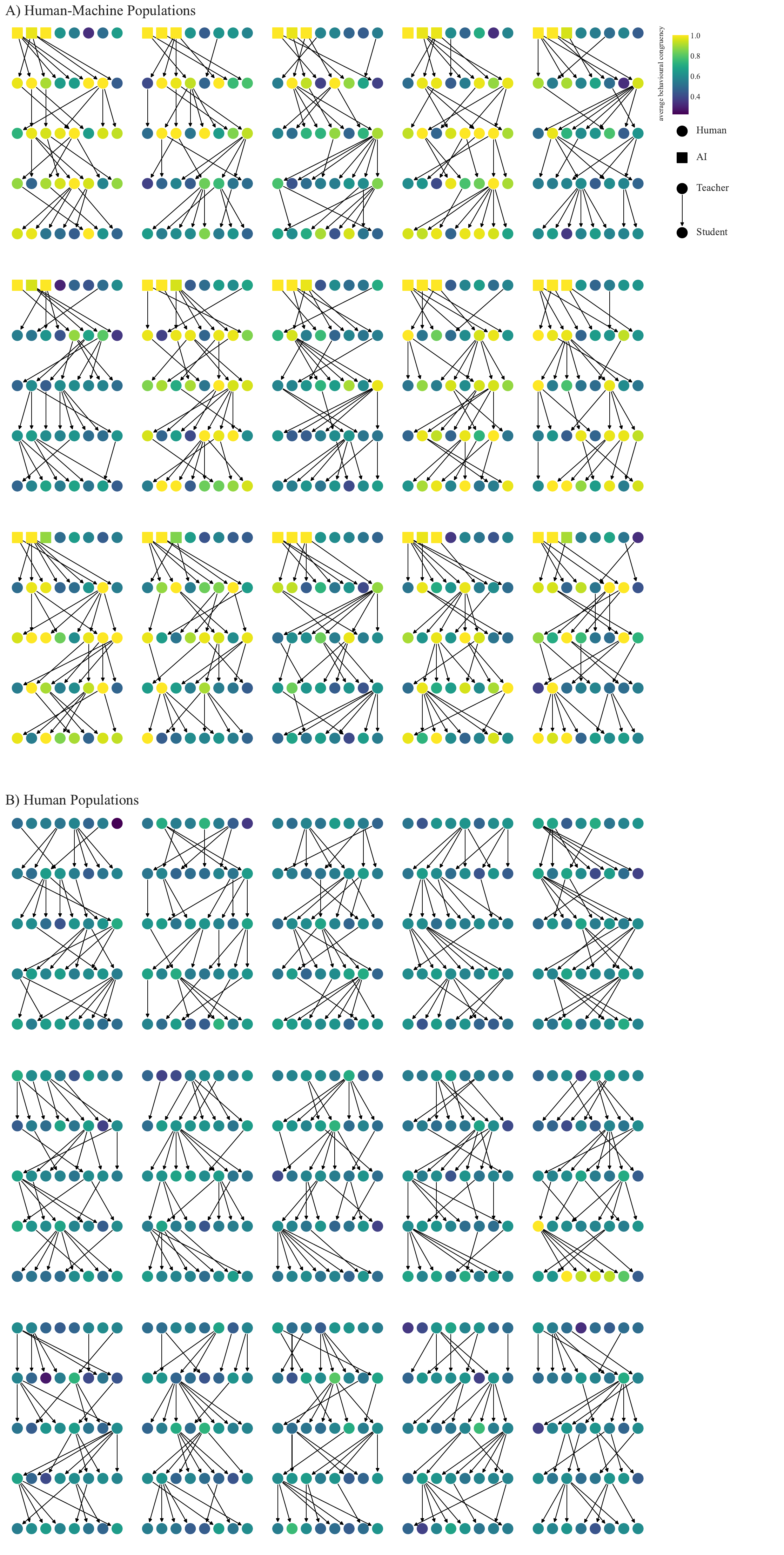}
\caption{\textbf{Behavioural Congruency of All 30 Populations}: Human players are represented by circles and machine players by squares, with the color of each symbol indicating the fraction of the player's moves that are congruent with those of the machine. The color scale ranges from 0 (light) to 1 (dark). A move is considered congruent if, given the same context (including historical actions), the machine would have made the same move. Arrows depict teacher-student relationships.}
  \label{fig:populations_alignment}
\end{figure}

\begin{figure}
  \centering
  \includegraphics[width=0.6\linewidth]{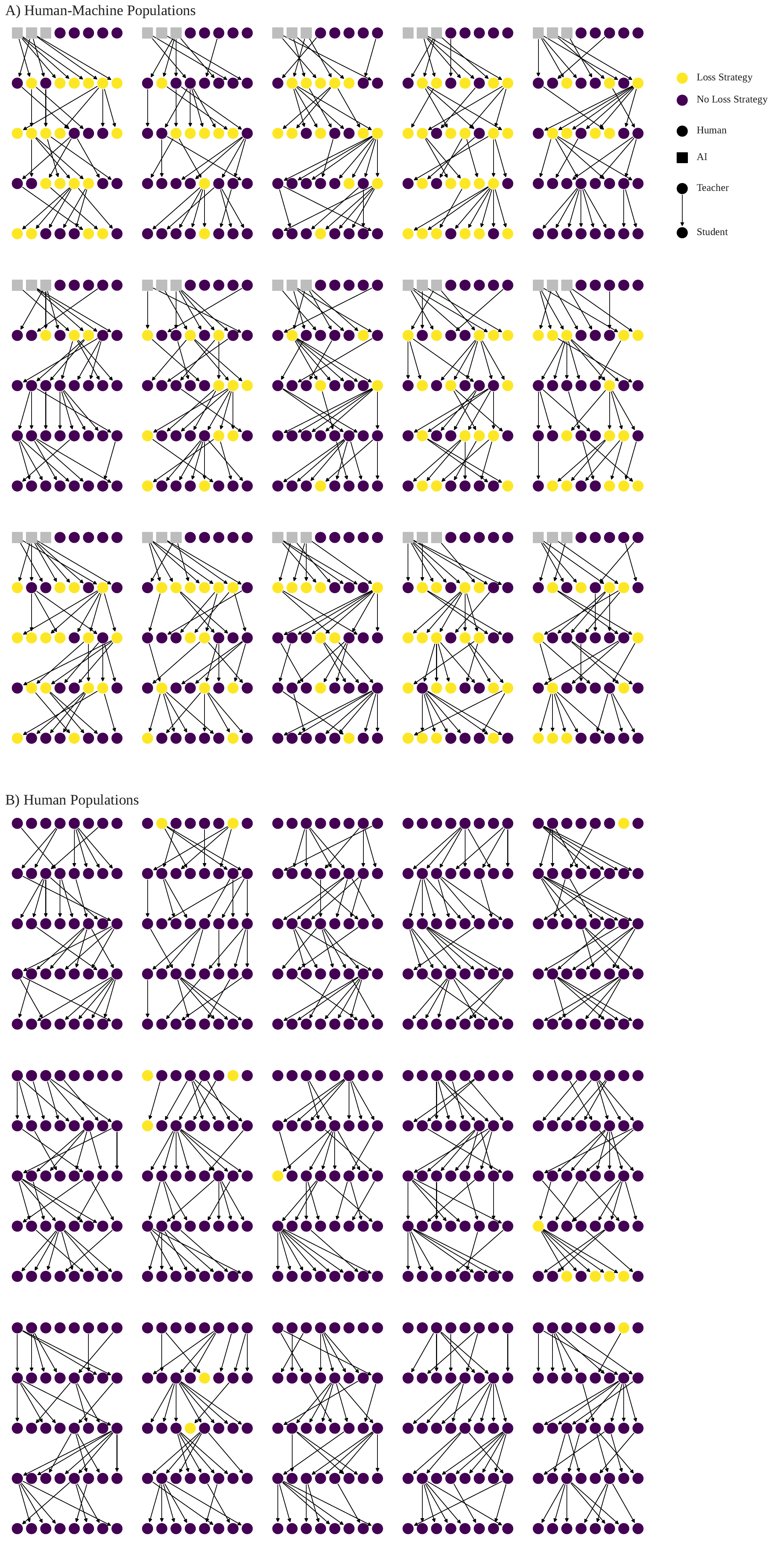}
    \caption{\textbf{Coded Strategies of All 30 Populations}: Human players are represented by circles and machine players by squares. A yellow circle signifies that the human player, after social learning, described a written strategy resembling deliberate acceptance of losses. Other strategies are represented by dark violet. Arrows depict teacher-student relationships.}
  \label{fig:populations_strategy}
\end{figure}

\begin{figure}
  \centering
  \includegraphics[width=1\linewidth]{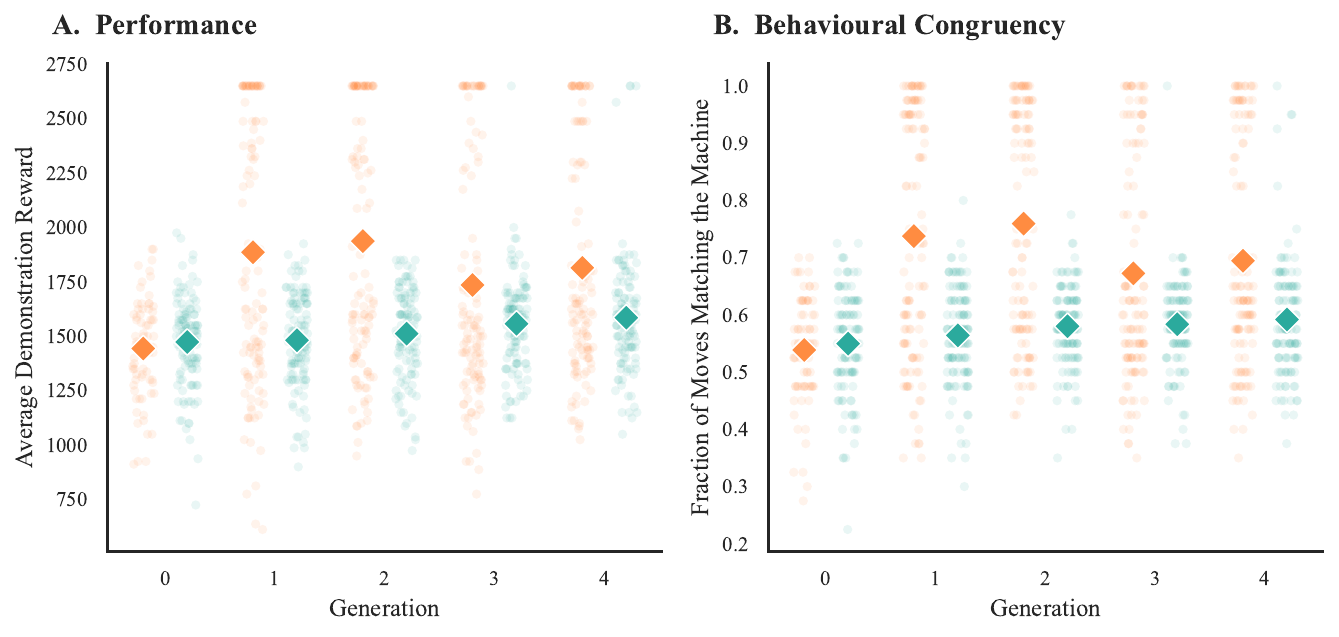}
\caption{\textbf{Individual Task Performance and Behavioural Congruency}:
Individual participants are represented by dots, and the grand mean for each condition is indicated by a diamond. Human-only populations are shown in green ($n = 600$ human participants across 15 populations), and human–machine populations in orange ($n = 555$ human participants across 15 populations). Performance (Panel A) refers to the average number of points earned per participant during the Demonstration Phase. Behavioural congruency (Panel B) is defined as the proportion of moves made by a human player that would also have been made by a machine player, conditional on identical preceding moves. Machine behaviour is operationalised using the majority vote across three independently trained models.}
  \label{fig:metrics_individual_level}
\end{figure}

\begin{figure} \centering \includegraphics[width=0.85\linewidth]{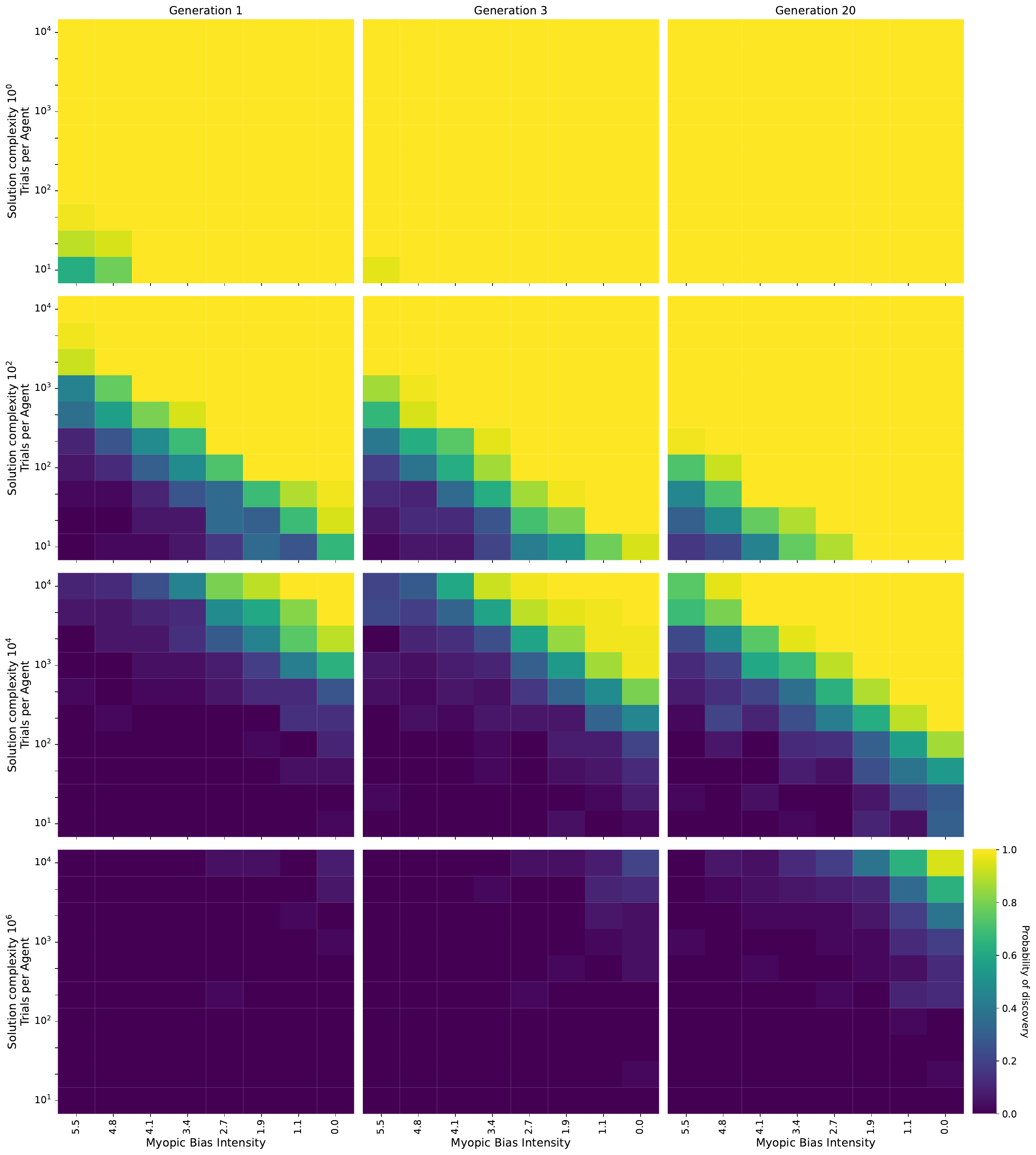} \caption{\textbf{Discovery probability as a function of exploration capacity and bias.} We simulated 20 agents over 20 generations, with each agent assigned a fixed number of trials per generation, as indicated by the x-axis of each subplot, and systematically varied the exploration bias toward myopic strategies, as indicated by the y-axis. Panels in different rows represent increasing solution complexity, while columns indicate the number of generations elapsed. Social learning parameters were held constant ($\tau = 0.1$ and $\lambda = 0.9$). Brighter colors indicate a higher probability that populations have discovered the optimal strategy. The heatmaps suggest that exploration capacity $K$ and bias $q$ constrain the discoverable solution complexity, with intergenerational transmission partially mitigating these constraints.}
\label{fig.bias_discoverability} 
\end{figure}

\begin{figure}
\centering
\includegraphics[width=1.0\linewidth]{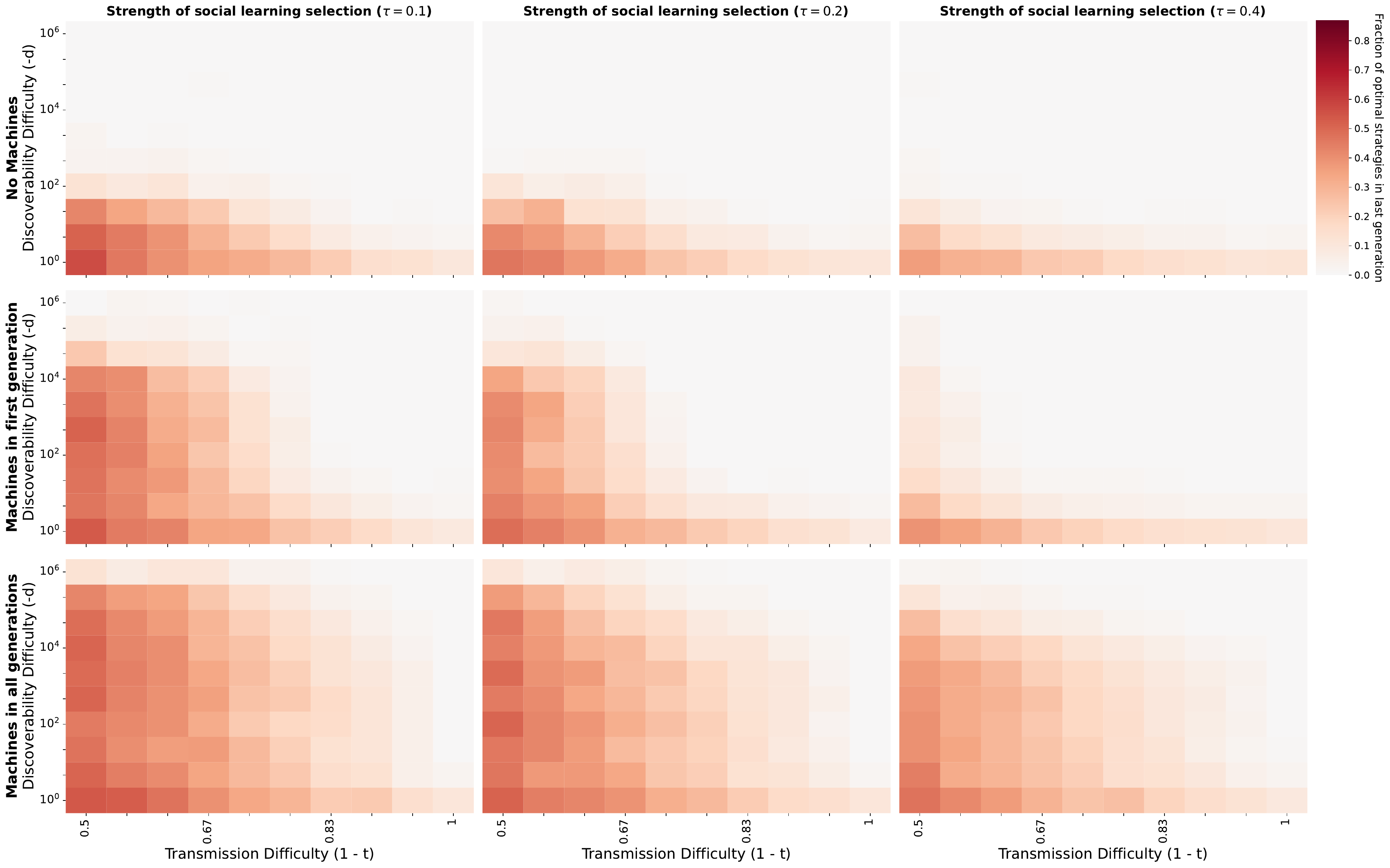}
\caption{\textbf{Selective social learning and machine availability.} We investigate how the strength of social learning and the availability of machines influence the adoption of the optimal solution. Social learning strength is operationalized via a softmax function over the performance of potential teachers, with lower temperature $\tau$ indicating stronger selection toward higher-performing teachers. We simulated 20 agents over 20 generations. We compare scenarios with no machines, five machines in the first generation only, and five machines present in all generations. Machine advantage and bias are set to $K_{\mathrm{machine}} = 10000$ and $q_{\mathrm{machine}} = 0.5$, respectively. Color indicates the fraction of human agents adopting the optimal strategy in the final generation, with the color scale ranging from 0 (light) to 1 (dark). Results suggest that stronger selective social learning enables populations to adopt strategies that are more difficult to discover and transmit. Selective social learning is particularly important for preserving high-performing, machine-discovered strategies across generations. By contrast, selective social learning is less consequential when machines are consistently present.}
\label{fig.abm_social_learning_selection}
\end{figure}

\begin{figure}
\centering
\includegraphics[width=1.0\linewidth]{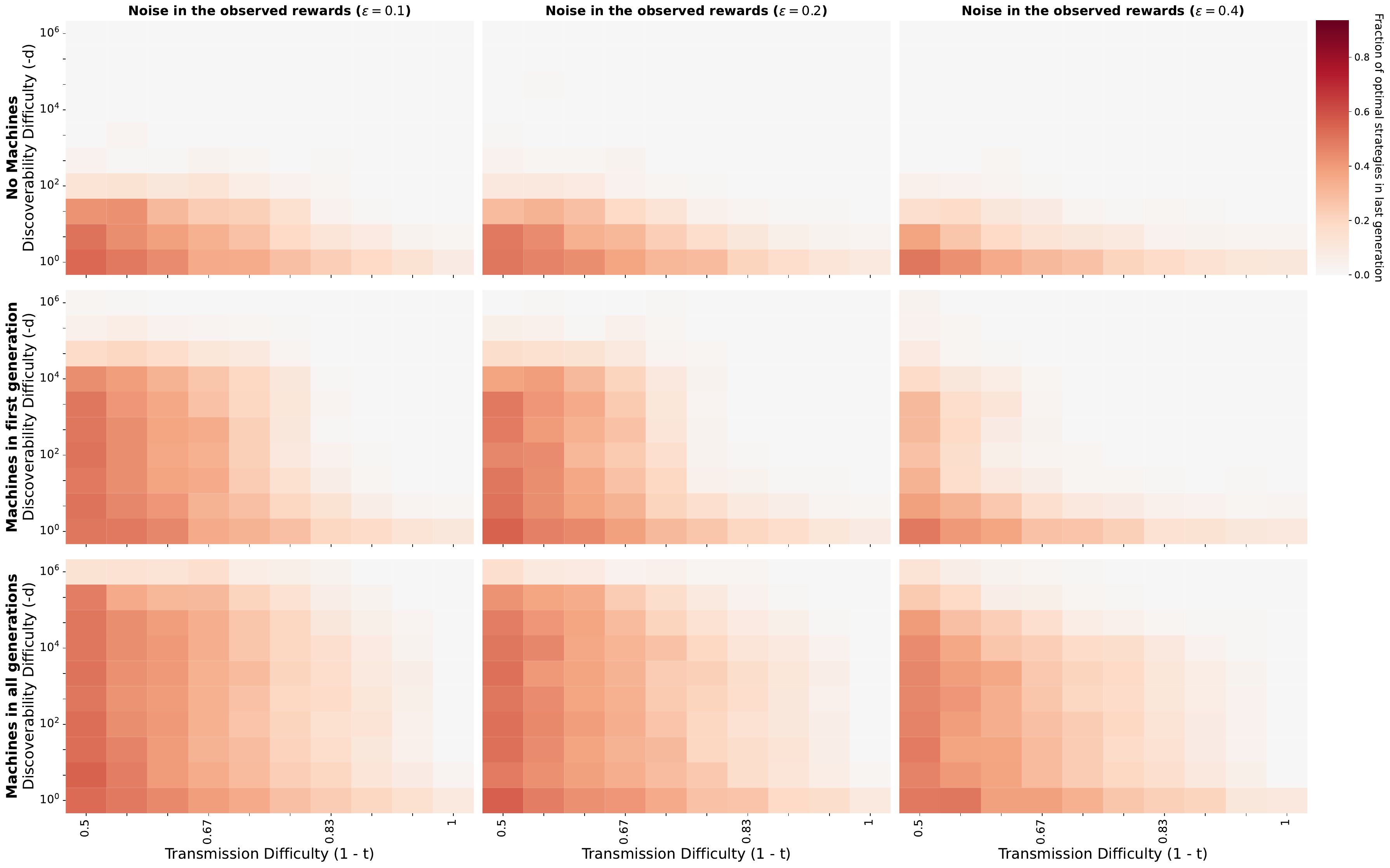}
\caption{\textbf{Selective social learning and machine availability.} We investigate how reward noise influences the adoption of the optimal solution under different machine-placement scenarios. Noise is introduced as independent Gaussian perturbations to individual realized rewards, with standard deviation $\epsilon \in {0.1, 0.2, 0.4}$. We simulate 20 agents over 20 generations and compare scenarios with no machines, five machines in the first generation only, and five machines present in all generations. Machine advantage and bias are set to $K_{\mathrm{machine}} = 10000$ and $q_{\mathrm{machine}} = 0.5$, respectively. Color indicates the fraction of human agents adopting the optimal strategy in the final generation, with the color scale ranging from 0 (light) to 1 (dark). Results indicate that increasing reward noise degrades the reliability of performance-based teacher selection, hindering the intergenerational transmission of difficult-to-discover strategies, particularly when machine contributions are transient.}
\label{fig.abm_reward_uncertainty}
\end{figure}

\begin{figure}
\centering
\includegraphics[width=0.7\linewidth]{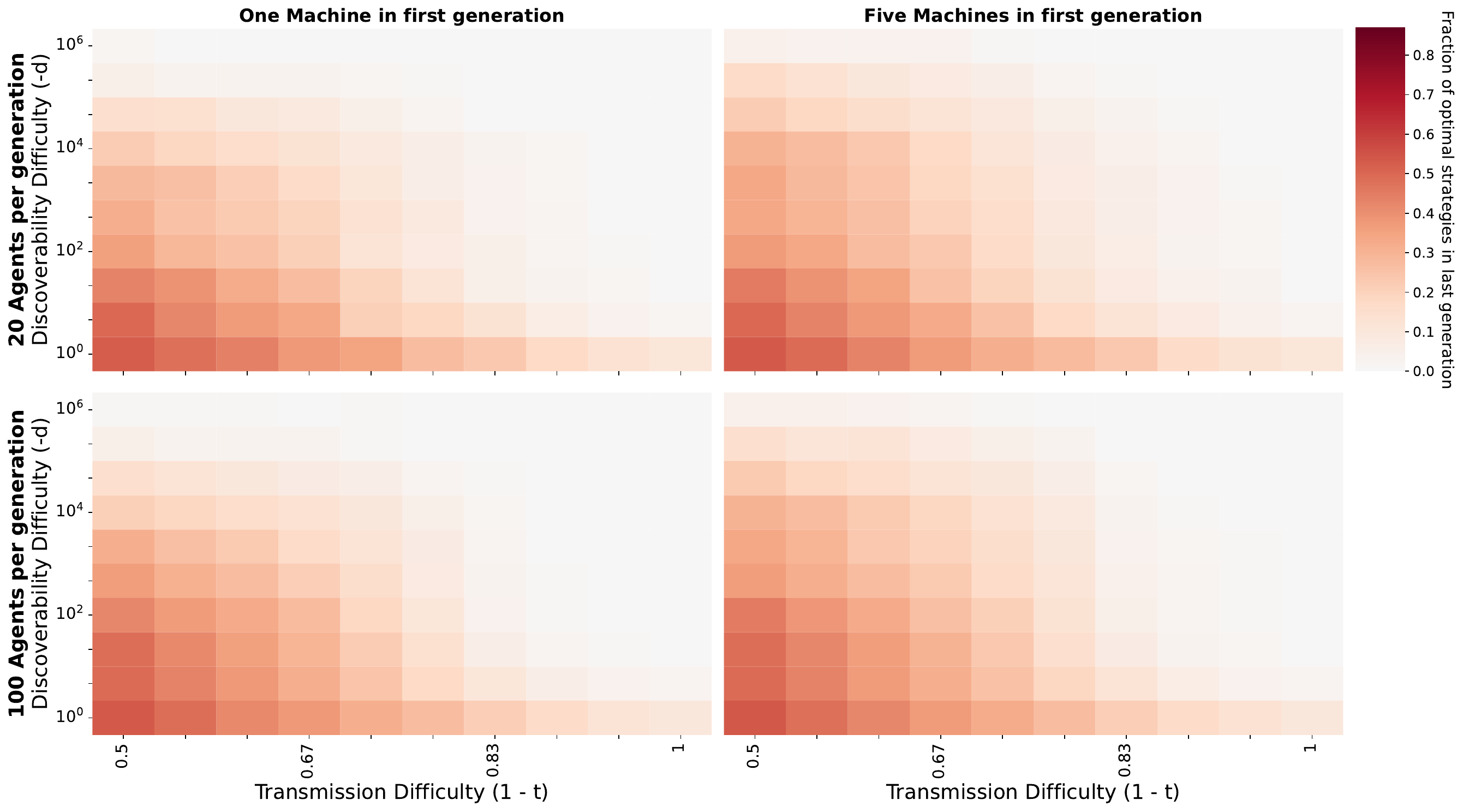}
\caption{\textbf{Robustness Analysis.} We examined the robustness of our results with respect to system size and the number of machines. Holding the social learning parameter ($\tau = 0.1$) and machine advantage ($K_{\mathrm{machine}} = 10000$ and $q_{\mathrm{machine}} = 0.5$) constant, we observed minimal sensitivity of the  to variation in both the number of machines in the first generation and the number of agents per generation. Color indicates the fraction of human agents adopting the optimal strategy in the final generation, with the color scale ranging from 0 (light) to 1 (dark).}
\label{fig.abm_robustness}
\end{figure}


\begin{table}
  \centering
  \label{tab:model_coeff_reward}
\begin{tabular}{@{\extracolsep{5pt}}lccc}
\\[-1.8ex]\hline
\hline \\[-1.8ex]
 & \multicolumn{3}{c}{Prediction} \\
\cline{2-4}
\\[-1.8ex] & \multicolumn{3}{c}{} \\
 & 1a & 1b & 2a \\
\hline \\[-1.8ex]
 Condition = Human Population & $-$309.2 & $-$229.1 & $-$404.6 \\
  & (57.0) & (82.4) & (59.4) \\
  & [$-418.4$, $-200.1$] & [$-383.1$, $-55.5$] & [$-524.8$, $-286.1$] \\
  & $P < 0.001$ & $P = 0.010$ & $P < 0.001$ \\
 Generation & $-$46.5 &  &  \\
  & (23.2) &  &  \\
  & [$-93.2$, $-2.7$] & & \\
  & $P = 0.054$ & & \\
 Condition:Generation & 86.4 &  &  \\
  & (32.8) &  &  \\
  & [$23.7$, $146.8$] & & \\
  & $P = 0.013$ & & \\
 Intercept & 1,843.5 & 1,814.8 & 1,886.5 \\
  & (40.3) & (58.3) & (42.0) \\
  & [$1763.5$, $1918.2$] & [$1698.1$, $1925.8$] & [$1806.6$, $1975.1$] \\
\hline \\[-1.8ex]
Observations & 3,840 & 960 & 960 \\
$R^2_{\text{marginal}}$ / $R^2_{\text{conditional}}$ & 0.078 / 0.529 & 0.045 / 0.509 & 0.110 / 0.571 \\
\hline
\hline \\[-1.8ex]
\end{tabular}
  \caption{Model coefficients predicting task performance. Estimated coefficients with standard errors (SE), 95\% bootstrap confidence intervals, and $P$-values (Satterthwaite approximation).}
\end{table} 


\begin{table}
  \centering
  \label{tab:model_coeff_congruency}
\begin{tabular}{@{\extracolsep{5pt}}lccc}
\\[-1.8ex]\hline
\hline \\[-1.8ex]
 & \multicolumn{3}{c}{Prediction} \\
\cline{2-4}
\\[-1.8ex] & \multicolumn{3}{c}{} \\
 & 1a & 1b & 2a \\
\hline \\[-1.8ex]
 Condition = Human Population
 & $-0.867$
 & $-0.644$
 & $-1.126$ \\
  & ($0.138$)
  & ($0.195$)
  & ($0.134$) \\
  &  [$-1.12$, $-0.61$]
     & [$-1.02$, $-0.25$]
     & [$-1.41$, $-0.88$] \\
  & $P < 0.001$
  & $P < 0.001$
  & $P < 0.001$ \\
 Generation
 & $-0.152$ &  &  \\
  & ($0.049$) &  &  \\
  &  [$-0.25$, $-0.05$] & & \\
  & $P = 0.002$ & & \\
 Condition:Generation
 & $0.201$ &  &  \\
  & ($0.068$) &  &  \\
  & [$0.07$, $0.33$] & & \\
  & $P = 0.003$ & & \\
 Intercept
 & $1.209$
 & $1.054$
 & $1.396$ \\
  & ($0.098$)
  & ($0.139$)
  & ($0.097$) \\
  & [$1.02$, $1.40$]
     & [$0.78$, $1.34$]
     & [$1.21$, $1.59$] \\
\hline \\[-1.8ex]
Observations
 & $38{,}400$
 & $9{,}600$
 & $9{,}600$ \\
$R^2_{\text{marginal}}$ / $R^2_{\text{conditional}}$
 & 0.047 / 0.227
 & 0.025 / 0.204
 & 0.070 / 0.272 \\
\hline
\hline \\[-1.8ex]
\end{tabular}
  \caption{Model coefficients predicting behavioural congruency. Estimated coefficients (log-odds) with standard errors (SE), 95\% bootstrap confidence intervals, and $P$-values (Wald $z$-test).}
\end{table}

\begin{table}
\centering
\begin{tabular}{@{\extracolsep{5pt}}lc}
\\[-1.8ex]\hline
\hline \\[-1.8ex]
 & \multicolumn{1}{c}{Prediction} \\
\cline{2-2}
\\[-1.8ex] &  \\
 & 2b \\
\hline \\[-1.8ex]
 Condition = Human Population & $-$4.180 \\
  & (0.617) \\
  & [$-6.30$, $-3.11$] \\
  & $P < 0.001$ \\
 Generation & $-$0.503 \\
  & (0.176) \\
  & [$-0.89$, $-0.10$]  \\
  & $P = 0.004$ \\
 Condition:Generation & 0.132 \\
  & (0.497) \\
  & [$-1.40$,$1.41$]  \\
  & $P = 0.609$ \\
 Intercept & $-$0.578 \\
  & (0.281) \\
  & [$-1.18$, $0.05$]  \\
\hline \\[-1.8ex]
Observations & 960 \\
$R^2_{\text{marginal}}$ / $R^2_{\text{conditional}}$ & 0.539 / 0.670 \\
\hline
\hline \\[-1.8ex]
\end{tabular}
  \caption{Model coefficients predicting written strategy. Estimated coefficients (log-odds) with standard errors (SE), 95\% bootstrap confidence intervals, and $P$-values (Wald $z$-test).}
\end{table}


\end{document}